\documentclass[11pt]{article}

\newlength{\boxparindent}
\usepackage{color}
\usepackage{graphics}
\usepackage{ifthen}
\usepackage{amsfonts}

\begin{document}

\definecolor{deepblue}{rgb}{0,0,0}
\definecolor{deepred}{rgb}{0,0,0}
\definecolor{red}{rgb}{0,0,0}
\definecolor{green}{rgb}{0,0,0}
\definecolor{blue}{rgb}{0,0,0}

\newcommand{\sect}[1]{section~\ref{#1}}
\newcommand{\se}[1]{\color{blue}
\section{\textsl{#1}} \color{black}}
\renewcommand{\thesection}{\textsl{\arabic{section}}}
\newcommand{\sse}[1]{\color{blue} \subsection{\textit{#1}} \color{black}}
\renewcommand{\thesubsection}{\textit{\arabic{section}.\arabic{subsection}}}
\newcommand{\ssse}[1]{\color{blue} \subsubsection{\small \textit{#1}}
\color{black}}
\renewcommand{\thesubsubsection}{\small
\textit{\arabic{section}.\arabic{subsection}.\arabic{subsubsection}}}
\newcommand{\be}{\color{deepred} \begin{equation}}
\newcommand{\ee}{\end{equation} \color{black}}
\newcommand{\ba}{\color{deepred} \begin{eqnarray}}
\newcommand{\ea}{\end{eqnarray} \color{black}}
\newcommand{\nn}{\nonumber}
\newcommand{\m}[1]{\textcolor{deepred}{$#1$}}
\renewcommand{\refname}{\textcolor{blue}{\textsl{References}}}
\renewcommand{\abstractname}{\large \textcolor{deepblue}{\textsf{Abstract}}}
\newcommand{\ca}[1]{\color{blue} \bfseries \slshape
\caption{#1} \color{black}}
\newcommand{\fig}[1]{figure~\ref{#1}}
\newcommand{\re}[1]{(\ref{#1})}
\newcommand{\maintitle}[1]{\vskip 0.9cm \begin{center}
{\LARGE{\bf \sffamily \textcolor{deepblue}{#1}}} \end{center} \vskip 0.4cm}
\newcounter{author}
\newcommand{\namecomma}[2]{\color{deepred}
\ifthenelse{\value{author}=99}{}{\begin{center} \setcounter{author}{99}} #1%
\footnote{E-mail: {\tt #2}},}
\newcommand{\name}[2]{\color{deepred}
\ifthenelse{\value{author}=99}{}{\begin{center} \setcounter{author}{99}} #1%
\footnote{E-mail: {\tt #2}}}
\newcommand{\andname}[2]{\color{deepred} and #1%
\footnote{E-mail: {\tt #2}}}
\newcommand{\address}[1]{\vskip 0.25cm
\centerline{\sl\small #1\,} \vskip 0.25cm
\end{center} \setcounter{author}{0}}
\newcommand{\lastaddress}[1]{\vskip 0.25cm
\centerline{\sl\small #1\,} \vskip 0.9cm
\end{center} \color{black} \setcounter{author}{0}}
\newcommand{\paperrefs}[1]{\color{deepred}
{\raggedleft #1 \\} \color{black} }
\newcommand{\rname}[1]{\newblock {\em #1,}}
\newcommand{\rauthor}[1]{#1,}
\newcommand{\journal}[4]{#1 \textbf{#2} (#3) #4}
\newcommand{\preprint}[1]{, preprint \texttt{#1}}
\newcommand{\xpreprint}[1]{ preprint \texttt{#1}}
\newcommand{\thline}{\hline\hline}

\newcommand{\la}{\lambda}
\newcommand{\xl}{\frac{\xi}{\lambda}}
\newcommand{\xp}{\frac{\xi}{\pi}}
\newcommand{\g}{\Gamma}
\newcommand{\ul}{\frac{\la u}{\pi}}
\newcommand{\hf}{\frac{1}{2}}
\newcommand{\hft}{{\textstyle\frac{1}{2}}}
\newcommand{\hp}{\frac{\pi}{2}}
\newcommand{\T}{\theta}
\newcommand{\pl}{\frac{\pi}{\lambda}}
\newcommand{\st}[1]{|#1\rangle}
\newcommand{\w}{\omega}
\newcommand\phup{^{\phantom{1}}}

\newtheorem{lemma}{Lemma}

\paperrefs{DTP-00/65 \\
\texttt{hep-th/0008071}}

\maintitle{Boundary Spectrum in the sine-Gordon Model with Dirichlet
Boundary Conditions}
\name{Peter Mattsson}{P.A.Mattsson@dur.ac.uk}
\andname{Patrick Dorey}{P.E.Dorey@dur.ac.uk}
\lastaddress{Dept.~of Mathematical Sciences,
University of Durham, Durham DH1 3LE, UK}

\begin{abstract} 
We find the spectrum of boundary bound states for the sine-Gordon
model with Dirichlet boundary conditions, closing the bootstrap and
providing a complete description of all the poles in the boundary
reflection factors. The boundary Coleman-Thun mechanism plays an
important role in the analysis. Two basic lemmas are introduced
which should hold for any 1+1-dimensional boundary field
theory, allowing the general method to be applied to other models.
\end{abstract}

\se{Introduction}
Integrable quantum field theories in two dimensions can be restricted
from the whole line by certain boundary conditions while still
preserving their integrability~\cite{GhoshZam}. As well as being of
theoretical interest, these models have interesting physical
applications, for example to impurity problems in an interacting 1D
electron gas~\cite{Kane} or edge excitations in fractional
quantum Hall states~\cite{Wen,Fendley}. (A recent review
can be found in \cite{Saleur}.)
  
One such theory is the sine-Gordon model, which has the bulk action
\be
\mathcal{A}_{SG}=\int_{-\infty}^{\infty}dx \int_{-\infty}^{\infty}dt~
\hft(\partial_{\mu}\varphi)^2
-\frac{m_0^{2}}{\beta^{2}}(\cos (\beta \varphi)-1)
\ee
on the full line, where \m{\varphi(x,t)} is a scalar field and
\m{\beta} is a dimensionless coupling constant. It was argued
in~\cite{GhoshZam} that this could be restricted to the half line \m{x
\in (-\infty,0]} while still preserving integrability by adding a
``boundary action'' term
\be
-\int_{-\infty}^{\infty}dt\, M \cos \left[\frac{\beta}{2}(\varphi\phup_B -
\varphi\phup_{0})\right],
\label{eq:boundact}
\ee    
where \m{M} and \m{\varphi_{0}} are free parameters,
and $\varphi\phup_B(t)=\varphi(x,t)|_{x{=}0}$. In addition to
the usual bulk bound states, in general this also presents us with 
a complicated spectrum of boundary bound states. This aspect was
mentioned briefly in~\cite{GhoshZam,Ghoshal}, and investigated in more
detail in~\cite{Skorik}. However in none of these works was a full
explanation of all poles achieved. In this paper we return to the
problem, and find that the full story appears to involve a much richer
structure of boundary bound states than has previously been suggested.
An important feature of our analysis, not taken into account in the
earlier works, is the so-called boundary Coleman-Thun
mechanism~\cite{DTW}. After a brief description of the boundary
sine-Gordon model in \sect{sec:review}, this mechanism is reviewed in
\sect{sec:colethun}. In the course of the section we also develop a
couple of useful lemmas, which simplify the subsequent discussions
considerably.

We will restrict ourselves to the case of Dirichlet boundary 
conditions, which simply fixes the value of the field at the boundary to
\m{\varphi(x=0,t)=\varphi_{0}} for all time. This corresponds to
taking \m{M \rightarrow \infty} above. After an initial discussion in 
\sect{sec:initpole}, in \sect{sec:example} we make a complete analysis
of a particular example. This serves to show many of the features of
the bound state structure, and is a useful warm-up to the discussion of the
general case in \sect{sec:general}. Finally, we gather together our
conclusions in \sect{sec:conclusions}. 

\se{The reflection factors}
\label{sec:review}
\sse{The theory in the bulk}
We begin with a summary of the theory in the bulk, mainly to set up
the notation. Further details can be found in the
review~\cite{Zsquare}. The bulk sine-Gordon model is known to be
integrable at both the classical and the quantum levels~\cite{Faddeev}. The
theory has an infinite number of degenerate vacua, with the discrete
symmetry \m{\varphi \rightarrow \varphi+\frac{2\pi}{\beta}m}, with
\m{m \in \mathbb{Z}}. The particle spectrum consists of a soliton
(\m{A}) and an anti--soliton (\m{\overline{A}}) --- both of which
interpolate between neighbouring vacua --- and a number of soliton --
anti-soliton bound states (``breathers'') \m{B_{n}},
\m{n=1,2,\ldots,<\la}, where
\be
\la = \frac{8\pi}{\beta^{2}}-1.
\ee  
The soliton (anti-soliton) has a topological charge of 1 (-1) while
the breathers are neutral. The soliton and anti-soliton both have the
same mass, which we shall call \m{m_{s}}, while the mass of
the $n^{\rm th}$ breather \m{B_{n}}
is \m{m_{n}=2m_{s}\sin \left(\frac{n\pi}{2\la}\right)}.

The integrability of the theory means that particle production is
forbidden, and all scattering is factorisable -- the amplitude for any
scattering process can be reduced to a product of two-particle
amplitudes. In addition, charge, parity and time-reversal (C, P and
T) symmetry holds for the bulk theory. If we denote the soliton S-matrix as
\m{S^{ab}_{cd}(\T)} for rapidity \m{\T}, with \m{a,b,c,d} taking the
value \m{+} (\m{-}) if the particle is a soliton (anti-soliton), the 
non-zero scattering amplitudes are \m{S^{++}_{++}(\T)=S^{--}_{--}(\T)=a(\T)}
(soliton-soliton or anti-soliton-anti-soliton scattering),
\m{S^{+-}_{+-}(\T)=S^{-+}_{-+}(\T)=b(\T)} (soliton-anti-soliton
transmission), and \m{S^{+-}_{-+}(\T)=S^{-+}_{+-}(\T)=c(\T)}
(soliton-anti-soliton reflection). Explicitly,
\ba
a(\T)&=&\sin [\la(\pi - u)]\rho(u)\,, \nonumber\\
b(\T)&=&\sin(\la u)\rho(u)\,, \\
c(\T)&=&\sin(\la \pi)\rho(u)\,, \nonumber
\ea
where \m{u=-i\T} and
\be
\rho(u)=\frac{1}{\sin (\la (u-\pi))}\prod_{l=1}^{\infty}\left[\frac{\g
\left((2l-2) \la -
\ul\right) \g \left(1+2l\la -\ul \right)}{\g \left( (2l-1)\la - \ul
\right) \g \left(1+ (2l-1)\la - \ul \right)} / (u \rightarrow - u) \right].
\ee
As pointed out in~\cite{Pillin},
this factor can also be written in terms of Barnes' 
diperiodic sine function \m{S_{2}(x|\omega_1,\omega_2)} \cite{Barnes,Jimbo}.
This is a meromorphic function parametrised by the pair of `quasiperiods'
\m{(\omega_1,\omega_2)}, with poles and zeroes at the following
points:
\ba
\hbox{poles}&:& x=n_1\,\omega_1+n_2\,\omega_2 ~~~~(n_1,n_2=1,2,\dots) \nn\\
\hbox{zeroes}&:& x=m_1\,\omega_1+m_2\,\omega_2~~~(m_1,m_2=0,-1,-2\dots)
\ea
In terms of this function,
\be
\rho(u)=\frac{1}{\sin (\la
(u-\pi))}\frac{S_{2}\left(\pi-u\left|\pl,2\pi\right.
\right)S_{2}\left(u\left|\pl,2\pi
\right. \right)}{S_{2}\left(\pi+u\left| \pl,2\pi
\right. \right)S_{2}\left( -u \left| \pl,2\pi \right. \right)}\,.
\ee

The amplitudes \m{b(\T)} and \m{c(\T)} have simple poles at \m{\T=i\left(\pi -
\frac{n\pi}{\la}\right)}, \m{n=1,2,\ldots,<\la}, which can be
attributed to the creation of \m{B_{n}} in the forward channel. There
are also poles at \m{\T=\frac{i\pi n}{\la}} in \m{a(\T)} and \m{b(\T)}
corresponding to the same process in the cross channel. 
Since all poles that we will be discussing, both in the bulk and at the
boundary, occur at imaginary rapidities, from now on we will
use the variable \m{u=-i\T} and always work in terms of
purely imaginary rapidities. 

\sse{Solitonic ground state reflection factors}
The general integral boundary condition found in \cite{GhoshZam} 
does not conserve topological charge, so in principle
four solitonic boundary reflection
factors need to be introduced, as well as a set of breather reflection
factors. The solitonic factors which we quote here were 
given in~\cite{GhoshZam}, while
breather factors can be found in~\cite{Ghoshal}. 
The reflection factors for the sine-Gordon solitons off the boundary
ground state will be denoted by
\m{P_{\pm}(u)} (a soliton or anti-soliton, incident on the boundary,
is reflected back unchanged) and \m{Q_{\pm}(u)} (a soliton is
reflected back as an anti-soliton, or vice versa). In the Dirichlet
case, topological charge is conserved and so \m{Q_{\pm}=0}. The
remaining factors can be written as\footnote{Note that there is a
small error in Ghoshal and Zamolodchikov's formula (5.23) for
\m{\sigma}, where the denominator should read
\m{\Pi(x,\pi/2)\Pi(-x,\pi/2)\Pi(x,-\pi/2)\Pi(-x,-\pi/2)}. We are
grateful to Subir Ghoshal for informing us of the corrected version.}
\be 
P^{\pm}(u)=R_{0}(u)
\prod_{l=1}^{\infty}\left[ \frac{\g \left(\hf + 2l\la \pm \xp+\ul
\right) \g \left( \hf + (2l-2)\la \mp \xp +\ul \right) }{\g \left( \hf
+ (2l-1)\la +\xp +\ul \right) \g \left( \hf + (2l-1)\la -\xp +\ul
\right)} /(u \rightarrow -u) \right]\,, 
\ee 
where 
\be
R_{0}(u)=\prod_{k=1}^{\infty}\left[\frac{\g \left( 1+\la(4k-4)
-\frac{2\la u}{\pi} \right)\g \left( 4\la k-\frac{2\la
u}{\pi}\right)}{\g \left( \la(4k-3)-\frac{2\la u}{\pi} \right) \g
\left( 1+ \la(4k-1)-\frac{2\la u}{\pi}\right)} /(u \rightarrow
-u)\right]\,,  
\label{eq:r0}
\ee
and \m{\xi = \frac{4\pi \varphi_{0}}{\beta}}. 
These factors can again be written in terms of Barnes' multiperiodic
functions, as
\be
P^{\pm}(u)=R_{0}(u)\frac{S_{2}\left(\frac{\pi}{2\la} \mp \xl+\pi+u \left|
\pl, 2\pi\right. \right)
S_{2}\left(\frac{\pi}{2\la} \mp \xl-u \left|
\frac{\pi}{\la},2\pi\right.\right)}
{S_{2}\left(\frac{\pi}{2\la} \mp \xl+\pi-u\left|\frac{\pi}{\la},2\pi\right.
\right)S_{2}\left(\frac{\pi}{2\la} \mp \xl+u\left|
\frac{\pi}{\la},2\pi\right. \right)},
\ee
with
\be
R_{0}(u)=\frac{S_{2}\left(\hp-u \left| \frac{\pi}{2\la},2\pi
\right. \right) S_{2}\left(\frac{\pi}{2\la}+u \left|
\frac{\pi}{2\la},2\pi \right. \right)}{S_{2}\left(\hp+u \left|
\frac{\pi}{2\la},2\pi \right. \right) S_{2}\left(\frac{\pi}{2\la}-u
\left| \frac{\pi}{2\la},2\pi \right. \right)}.
\ee
The theory is invariant under 
\m{\varphi_0\to\varphi_0+\frac{2\pi}{\beta}}, and also
under the simultaneous transformations \m{\varphi_0 \rightarrow
-\varphi_0} and
\m{\mathrm{soliton\ } \rightarrow \mathrm{anti{-}soliton}}. 
Introducing the boundary breaks the degeneracy of the bulk vacua, and
selects the lower line in \fig{fig:vacua} as the lowest-energy state,
with the upper line as the first excited state. Continuing
\m{\varphi_{0}} through \m{\frac{\pi}{\beta}} thus simply interchanges
the r\^{o}les of these two states, and selects the upper one as the
ground state.

\begin{figure}
\begin{center}
\unitlength 1.00mm
\linethickness{0.4pt}
\begin{picture}(50.11,55.00)(0,5)
\put(46.00,3.00){\rule{2.00\unitlength}{52.00\unitlength}}
\color{green}
\qbezier(45.39,29.44)(41.28,23.33)(32.61,22.56)
\qbezier(45.39,29.56)(44.50,39.89)(32.61,40.89)
\put(32.61,40.89){\line(-1,0){27.17}}
\put(32.61,22.56){\line(-1,0){27.17}}
\color{blue}
\multiput(5.44,41.00)(1,0){41}{\circle*{0.25}}
\multiput(5.44,22.22)(1,0){41}{\circle*{0.25}}
\color{deepred}
\put(3.11,9.11){\vector(0,1){6.67}}
\put(3.11,9.11){\vector(1,0){7}}
\put(3.11,17.67){\makebox(0,0)[cc]{$\varphi$}}
\put(11.89,9.11){\makebox(0,0)[cc]{$x$}}
\put(50.11,22.22){\makebox(0,0)[cc]{$0$}}
\put(50.11,41.00){\makebox(0,0)[cc]{$\frac{2\pi}{\beta}$}}
\put(50.61,29.50){\makebox(0,0)[cc]{$\varphi_{0}$}}
\color{black}
\end{picture}
\end{center}
\ca{Vacuum structure}
\label{fig:vacua}
\end{figure}
     
In light of this, we are free to 
choose \m{\xi}
to be in the interval
\be
0<\xi<\frac{4\pi^{2}}{\beta^{2}}=\frac{\pi(\la+1)}{2}.
\ee
Note that the topological
charge of the ground state is no longer zero, as in the bulk model,
but
\be
q=\frac{\beta}{2\pi}\int_{-\infty}^{0}dx\frac{\partial}{\partial
x}\varphi (x,t)=\frac{\beta}{2\pi}[\varphi (0,t)-\varphi
(-\infty,t)]=\frac{\beta \varphi_{0}}{2\pi}\,,
\ee
with the charge of the first excited state being \m{1-\frac{\beta
\varphi_{0}}{2\pi}}. We will find that all the boundary states have
one of these charges so, for convenience, we shall designate them
simply as 0 and 1 respectively.

\sse{Breather 
ground state
reflection factors}
For the breather sector, Ghoshal~\cite{Ghoshal} 
obtained the
relevant reflection factors --- \m{R^{n}_{\st{0}}(u)} for breather \m{n}
and boundary ground state \m{\st{0}} ---
from the solitonic
reflection factors using the general boundary bootstrap equation
\cite{FringK,GhoshZam}
\be
f^{n}_{i_{1}i_{2}}R^{i_{1}}_{j_{1}\st{x}}\left(u+\frac{u_{n}}{2}\right)
S_{j_{2}f_{1}}^{i_{2}j_{1}}(2u)R^{j_{2}}_{f_{2}\st{x}}\left(u-\frac{u_{n}}{2}
\right)=f^{n}_{f_{1}f_{2}}R^{n}_{\st{x}}(u),
\ee
where \m{u_{n}=\pi - \frac{n\pi}{\la}}, and the \m{R^{a}_{b\st{x}}(u)} are
the solitonic reflection factors, such that \m{R^{+}_{-\st{x}}(u)} is the
factor for a soliton to be reflected back as an anti-soliton and so
on. The \m{f^{n}_{ab}} are the bulk vertices for the creation of
breather \m{n} from (anti-)solitons \m{a} and \m{b}. 
These obey \m{f^{n}_{+-}=(-1)^{n}
f^{n}_{-+}}. The bootstrap is illustrated in \fig{fig:brboot}. 

\begin{figure}
\parbox{2.8in}{
\unitlength 1.00mm
\linethickness{0.4pt}
\begin{picture}(48.00,55.00)
\color{blue}
\multiput(35.00,44.89)(-1,1){10}{\circle*{0.25}}
\color{green}
\put(35.11,44.89){\line(2,-1){10.78}}
\put(45.89,39.22){\line(-2,-1){31.33}}
\put(45.89,23.22){\line(-1,-2){9.61}}
\put(35.22,44.67){\line(1,-2){10.67}}
\color{deepred}
\put(12.22,22.33){\makebox(0,0)[cc]{$f_{1}$}}
\put(35.11,2.78){\makebox(0,0)[cc]{$f_{2}$}}
\put(33.61,42.61){\makebox(0,0)[cc]{$i_{2}$}}
\put(37.94,45.72){\makebox(0,0)[cc]{$i_{1}$}}
\put(41.11,28.67){\makebox(0,0)[cc]{$j_{2}$}}
\put(43.78,36.17){\makebox(0,0)[cc]{$j_{1}$}}
\qbezier(41.67,32.11)(43.78,33.89)(46.00,32.56)
\put(51.00,30.94){\vector(-1,0){7}}
\put(52.00,30.94){\makebox(0,0)[lc]{$u-\frac{u_{n}}{2}$}}
\qbezier(39.78,42.44)(42.00,45.00)(46.00,44.44)
\put(51.00,42.56){\vector(-1,0){7}}
\put(52.00,42.56){\makebox(0,0)[lc]{$u+\frac{u_{n}}{2}$}}
\qbezier(32.89,47.11)(38.89,50.78)(45.89,49.00)
\put(39.11,50.78){\makebox(0,0)[cc]{$u$}}
\qbezier(37.33,40.56)(39.56,40.33)(40.00,42.33)
\put(33.44,38.78){\vector(4,3){4.11}}
\put(31.67,37.67){\makebox(0,0)[cc]{$u_{n}$}}
\color{black}
\put(46.00,3.00){\rule{2.00\unitlength}{52.00\unitlength}}
\end{picture}
} \ \LARGE = \normalsize \
\parbox{2.8in}{
\unitlength 1.00mm
\linethickness{0.4pt}
\begin{picture}(48.00,55.00)
\put(46.00,3.00){\rule{2.00\unitlength}{52.00\unitlength}}
\color{green}
\put(16.56,11.89){\line(2,1){13.22}}
\put(29.78,18.50){\line(-1,-3){4.20}}
\color{blue}
\multiput(29.67,18.44)(1,1){17}{\circle*{0.25}}
\multiput(29.67,51.11)(1,-1){17}{\circle*{0.25}}
\color{deepred}
\qbezier(42.00,38.89)(43.56,40.56)(46.00,39.56)
\put(43.78,41.44){\makebox(0,0)[cc]{$u$}}
\qbezier(24.67,15.89)(25.22,13.56)(28.11,13.33)
\put(24.44,13.00){\makebox(0,0)[cc]{$u_{n}$}}
\put(14.33,10.89){\makebox(0,0)[cc]{$f_{2}$}}
\put(24.22,4.33){\makebox(0,0)[cc]{$f_{1}$}}
\color{black}
\end{picture}
}
\ca{Breather bootstrap}
\label{fig:brboot}
\end{figure}

In the Dirichlet case, with topological
charge conserved, the bootstrap equation reduces to
\be
f^{n}_{i_{1}i_{2}}P^{i_{1}}_{\st{x}}\left(u+\frac{u_{n}}{2}\right)S^{i_{2}
i_{1}}_{f_{2}f_{1}}(2u)P^{f_{2}}_{\st{x}}\left(u-\frac{u_{n}}{2}\right)=
f^{n}_{f_{1}f_{2}}R^{n}_{\st{x}}(u). 
\ee 

Ghoshal found that, for the boundary ground state, the breather
reflection factors were
\be
R^{n}_{\st{0}}(u)=R^{(n)}_{0}(u)R^{(n)}_{1}(u),
\label{gsrfl}
\ee
where
\be 
R_{0}^{(n)}(u)=\frac{\left(\hf\right)
\left(\frac{n}{2\la}+1\right)}{\left( \frac{n}{2\la} +
\frac{3}{2}\right)} \prod_{l=1}^{n-1}\frac{ \left( \frac{l}{2\la}
\right) \left( \frac{l}{2\la} +1 \right)}{ \left(
\frac{l}{2\la}+\frac{3}{2}\right)^{2}}, 
\ee 
and 
\be
R_{1}^{(n)}(u)=\prod_{l=\frac{1-n}{2}}^{\frac{n-1}{2}}\frac{\left(
\frac{\xi}{\la \pi}-\hf +\frac{l}{\la}\right)}{\left( \frac{\xi}{\la
\pi} +\hf + \frac{l}{\la}\right)}.  
\ee
This makes use of the notation
\be
(x)=\frac{\sinh \left( \frac{\T}{2}+\frac{i\pi x}{2}\right)}{
\sinh \left( \frac{\T}{2}-\frac{i\pi x}{2} \right)},
\ee
which will also be helpful later.

\se{The boundary Coleman-Thun mechanism}
\label{sec:colethun}
In this section we will discuss some general features of the pole analysis of
boundary reflection factors, in preparation for the specific case of the
sine-Gordon model. The main
aim is to establish a couple of lemmas which will simplify the
subsequent discussion. We begin by recalling
the story in the bulk.

An initially mysterious feature of the sine-Gordon S-matrix
was the presence of a number of double
poles in the breather scattering amplitudes. 
At first it was thought that these might be related to the
integrability of the model, and it was only with the work of Coleman and
Thun~\cite{CT} that it was realised that they had a `prosaic'
origin as anomalous threshold poles, and could be explained using
standard field-theoretical ideas. Studies of affine Toda field 
theories~\cite{CM,BCDS,BCDSpole,del}
returned to this topic, and it proved possible, in certain cases,
to confirm the scenario of Coleman and Thun through standard, albeit
elaborate, perturbative calculations~\cite{BCDSpole}.

One element of Coleman and Thun's analysis 
was the observation that in 1+1 dimensions sometimes even
simple poles have 
complicated explanations, as anomalous thresholds.
This breaks
the usual association of every simple S-matrix 
pole with a bound state
in either the forward or the crossed channel.
The same mechanism is at work in the S-matrices of non self-dual
affine Toda field theories \cite{del}, as
was pointed out in~\cite{CDS}. 
This material is reviewed at greater length in~\cite{pedrev};
here, we are more interested in what can occur when a boundary is also
involved. This was discussed in~\cite{DTW} via a particularly simple
example, the boundary scaling Lee-Yang model, for which the spectrum of
boundary bound states had previously been found by other
means~\cite{DPTW}. 
It was found that the Coleman-Thun mechanism did indeed play a role,
there being a number of simple poles in the reflection
factors at locations which did not correspond to boundary bound states, but
which rather could be explained through on-shell (anomalous threshold)
diagrams for multiple rescattering processes involving the boundary.
Further work, applying the
method to affine Toda field theories, can be found 
in~\cite{DG}. 
It is important to remark that, both in the bulk and at the boundary, 
an anomalous threshold diagram would normally lead to a pole of order
greater than one. To match a simple pole, 
this order must be reduced somehow, and in
\cite{DTW,DG} this was always achieved by one or more
\lq internal' reflection factors having zeroes exactly at the 
point where the diagram went on-shell. 
Here, we will see both this mechanism of order reduction
(which mimics
that seen in bulk affine Toda theories \cite{CDS}) and cancellations
between different diagrams, closer to the original situation discussed by 
Coleman and Thun~\cite{CT}.

One difficulty, especially serious in cases when the spectrum of boundary
bound states is not known a priori, is the greatly increased complexity
of the on-shell diagrams once a boundary is involved. This makes it hard
to be sure that any given pole really does correspond to a new boundary
bound state.
In the bulk, a simple geometrical argument shows 
that poles in the S-matrix elements of
the lightest particle can never be explained by a Coleman-Thun
mechanism, and so must always be due to bound states~\cite{BCDS}. 
We wish to find analogous criteria for the boundary
situation. To this end, the following two lemmas turn out to be useful.
Suppose the incoming particle is of type \m{a}, and that its
reflection factor has a simple pole at \m{\T=iu}. 

\begin{lemma}
Let \m{\overline{U}_{a}=\min_{b,c} \left(\pi-U^{c}_{ab}\right)}. 
If \m{u<\overline{U}_{a}}, then the 
the pole at $iu$ cannot be explained by a Coleman-Thun mechanism, and so
must correspond to the binding of particle $a$ to the
boundary, either before or after crossing the outgoing particle.
\label{lemma:1}
\end{lemma}
{\bf Proof: }%
All processes must take the form shown in \fig{fig:lemma1} or the
crossed version shown in \fig{fig:lemma1c}.  Conservation of momentum
demands that all 
rescattering
must take place within the hatched region,
which is drawn 
from
the furthest point from the boundary where either
the incoming or outgoing particle undergoes any interaction.
If neither particle decays, we simply have a diagram of the form of
\fig{fig:sboun} or \fig{fig:uboun}. Otherwise, 
momentum conservation requires
that neither product of the particle which decays on the
boundary of the hatched region has a trajectory which takes it outside
that region. Fixing the notation by \fig{fig:decay} (with
angles \m{U^{b}_{ac}} and \m{U^{c}_{ab}} defined
correspondingly), this reduces to demanding \m{\pi -
U^{c}_{ab}\leq u \leq U^{b}_{ac}}.
If we introduce
\m{\overline{U}_{a}} then we must have \m{\overline{U}_{a} \leq u \leq
\pi - \overline{U}_{a}} (i.e. just \m{u \geq \overline{U}_{a}}, as
\m{u \leq \hp}). Thus, if \m{u < \overline{U}_{a}}, then the only
possible explanations for the pole are \fig{fig:sboun} and
\fig{fig:uboun}.

\begin{lemma}
If the boundary is in its ground state, then lemma~\ref{lemma:1} can
be strengthened, requiring that the incoming particle bind to the
boundary if \m{u} is outside the range
\m{\overline{U}_{a}<u<\hp-\overline{U}_{a}}. In addition, if
\m{\min_{b,c}U^{a}_{bc}>\hp}, the incoming particle must always bind
to the boundary.
\label{lemma:2}
\end{lemma}
{\bf Proof: }%
With the boundary in its ground state,
all rescattering must take
place in the area shown in \fig{fig:lemma2}. Reasoning as before but
demanding that both product particles be emitted into this more
restricted region, we find \m{\pi - U^{c}_{ab}\leq u \leq
U^{b}_{ac}-\frac{\pi}{2}}, or \m{\overline{U}_{a} \leq u
\leq \frac{\pi}{2}-\overline{U}_{a}}. 
In addition, both particles \m{b,c} must be 
emitted into an angle of \m{\hp}, so \m{U^{a}_{bc}<\hp} for at least
one pair of particles $b$, $c$.
If either of these conditions are violated,
then the incoming particle must bind to the boundary.

\begin{figure}
\begin{center}
\parbox{3in}{
\centering 
\unitlength 1.00mm
\linethickness{0.4pt}
\begin{picture}(42.00,55.00)(6,0)
\color{deepred}
\multiput(26.22,55.00)(0,-1){52}{\line(0,-1){0.5}}
\multiput(26.22,55.00)(0,-4){13}{\line(6,-1){19.78}}
\multiput(46.00,55.00)(0,-4){13}{\line(-6,-1){19.78}}
\color{green}
\put(6.11,5.33){\line(3,2){20.11}}
\put(6.11,52.56){\line(3,-2){20.11}}
\color{black}
\put(46.00,3.00){\rule{2.00\unitlength}{52.00\unitlength}}
\end{picture}
\ca{General process, uncrossed}
\label{fig:lemma1} } \
\parbox{3in}{ 
\centering
\unitlength 1.00mm
\linethickness{0.4pt}
\begin{picture}(42.00,55.00)(6,0)
\color{deepred}
\multiput(26.22,55.00)(0,-1){52}{\line(0,-1){0.5}}
\multiput(26.22,55.00)(0,-4){13}{\line(6,-1){19.78}}
\multiput(46.00,55.00)(0,-4){13}{\line(-6,-1){19.78}}
\color{green}
\put(26.22,18.74){\line(-3,2){20.11}}
\put(26.22,39.15){\line(-3,-2){20.11}}
\color{black}
\put(46.00,3.00){\rule{2.00\unitlength}{52.00\unitlength}}
\end{picture}
\ca{General process, crossed}
\label{fig:lemma1c} }
\end{center}
\end{figure}

\begin{figure}
\begin{center}
\parbox{3in}{ 
\centering
\unitlength 1.00mm
\linethickness{0.4pt}
\begin{picture}(42.00,55.00)(6,0)
\color{green}
\put(6.11,5.33){\line(3,2){20.11}}
\put(6.11,52.56){\line(3,-2){20.11}}
\color{deepred}
\put(26.22,18.67){\dashbox{0.5}(19.89,20.44)[cc]{}}
\multiput(26.22,38.61)(0,-4){5}{\line(6,-1){19.78}}
\multiput(46.00,38.61)(0,-4){5}{\line(-6,-1){19.78}}
\color{black}
\put(46.00,3.00){\rule{2.00\unitlength}{52.00\unitlength}}
\end{picture}
\ca{General ground state process}
\label{fig:lemma2} } \
\parbox{3in}{ 
\centering
\unitlength 1.00mm
\linethickness{0.4pt}
\begin{picture}(34.00,55.00)(9,0)
\color{green}
\put(27.92,28.31){\line(1,1){15.44}}
\put(27.81,28.31){\line(1,-1){15.44}}
\color{blue}
\put(27.89,28.44){\line(-1,0){19}}
\color{deepred}
\qbezier(32.31,32.53)(34.64,28.42)(32.09,24.19)
\put(40.41,28.31){\makebox(0,0)[cc]{$U^{a}_{bc}$}}
\put(18.31,26.19){\makebox(0,0)[cc]{$a$}}
\put(35.03,38.61){\makebox(0,0)[cc]{$b$}}
\put(35.03,17.59){\makebox(0,0)[cc]{$c$}}
\color{black}
\end{picture}
\ca{Decay process}
\label{fig:decay} 
}
\end{center}
\end{figure}

\vspace{3mm}

These two results, between them, will allow the spectrum of the 
sine-Gordon model with Dirichlet boundary
conditions to be fixed
completely, provided it is assumed that no pole corresponds to the
creation of a boundary state if it has an alternative (Coleman-Thun)
explanation. 

For the problem under discussion, writing the 
rapidity bounds \m{\overline{U}_a}
as \m{\overline{U}_{+(-)}} for the soliton (anti-soliton)
and 
as \m{\overline{U}_{n}} for the \m{B_{n}}, we have
\ba
\overline{U}_{\pm} &=& \frac{\pi}{2}-\frac{n_{\mathrm{max}}\pi}{2\la}
\nonumber \\
\overline{U}_{n}&=&\frac{\pi}{2\la}~,~~~ n \neq n_{max} \\
\overline{U}_{n_{\mathrm{max}}}&=&\hp-\frac{n_{\mathrm{max}}\pi}{2\la}\,,
\nonumber 
\ea
where \m{B_{n_{max}}} is the highest-numbered breather present in the
model. To derive these results, note that a soliton (anti-soliton)
can only decay into an anti-soliton (soliton) and a breather (with
vertex \m{U^{\pm}_{\mp n}=\hp+\frac{n\pi}{2\la}}). A breather can
either decay into a soliton--anti-soliton pair
(\m{U^{n}_{+-}=\pi-\frac{n\pi}{\la}}) or a pair of breathers
(\m{U^{l}_{nm}=\pi - \frac{l\pi}{2\la}} with \m{n=m+l} or \m{m=n+l}, or 
\m{U^{l}_{nm}=\frac{\pi(n+m)}{2\la}} with \m{l=n+m}).

These restrictions can also be combined to produce a stronger version of
lemma 1 when the incoming particle is a soliton.
If \m{\overline{U}_{+}<u<\frac{\pi}{\la}}, decay within the hatched
region is only possible into
the topmost breather and an anti-soliton. One or other of these particles
will be heading away from the centre of the diagram. If the process in
uncrossed, as in
\fig{fig:lemma1}, the breather will be created heading towards
the centre of the diagram, the anti-soliton away (we are being
somewhat cavalier with the direction of time; this should be
considered as a purely geometric argument). The anti-soliton must itself
obey our lemmas; if in any further  decay before it reaches the boundary
one of the decay products is heading away from the boundary, then there
would be no way to close the diagram while conserving momentum at every
vertex. For a crossed process (\fig{fig:lemma1c}) the breather is
the outermost particle, and is again restricted in its decay by
our lemmas for the same reason.

The anti-soliton created by the uncrossed process
heads for the boundary with a rapidity less than
\m{\overline{U}_{-}} and so, by
lemma~\ref{lemma:1}, may not decay. By the same token, the breather of
the crossed process cannot decay either so, 
if the initial soliton is not to form a bound state, the only possible
alternative processes 
are \fig{fig:fb3} and \fig{fig:fb2}. If these are found not
to occur (for example, if the necessary boundary vertices are not
present) 
then the pole must correspond to a bound state for any \m{u<\frac{\pi}{\la}}.

\se{Initial pole analysis}
\label{sec:initpole}
\sse{Solitonic ground state factors}
The \m{R_{0}(u)} factor is insensitive to the boundary parameters, and so all
its poles should be explicable in terms of the bulk. The only 
poles are at \m{u=\frac{N\pi}{2\la}}, where
\m{N=1,2,3,\ldots}, with no zeroes. These can be explained by the
creation of a breather which is incident perpendicularly on the boundary,
as shown in \fig{fig:bbre}. Here, 
as in all subsequent diagrams, the time axis points up the page,
and the \m{x} axis points to the right. Solitons and anti-solitons
are drawn as solid lines, while breathers are drawn as dotted
lines.

\begin{figure}
\begin{center}
\parbox{2.2in}{
\centering
\unitlength 1.00mm
\linethickness{0.4pt}
\begin{picture}(35.5,55.00)(12.5,0)
\put(46.00,3.00){\rule{2.00\unitlength}{52.00\unitlength}}
\color{green}
\put(27.92,28.31){\line(-1,1){15.44}}
\put(27.81,28.31){\line(-1,-1){15.44}}
\color{blue}
\multiput(27.89,28.44)(1,0){5}{\circle*{0.25}}
\color{deepred}
\qbezier(23.47,32.53)(21.14,28.42)(23.69,24.19)
\put(15.37,28.31){\makebox(0,0)[cc]{$\pi-\frac{N\pi}{\lambda}$}}
\put(37.47,26.19){\makebox(0,0)[cc]{$B_{N}$}}
\color{blue}
\multiput(33.00,28.44)(1,0){14}{\circle*{0.25}}
\color{black}
\end{picture}
\ca{$\mathbf{\xi}$-independent pole} 
\label{fig:bbre} }
\parbox{2.2in}{
\centering
\unitlength 1.00mm
\linethickness{0.4pt}
\begin{picture}(17.75,55.00)(30.25,0)
\put(46.00,3.00){\rule{2.00\unitlength}{52.00\unitlength}}
\color{green}
\put(30.25,5.53){\line(1,1){15.56}}
\put(45.69,37.08){\line(-1,1){15.56}}
\color{red}
\multiput(45.78,21.44)(0,1){16}{\circle*{0.25}}
\color{black}
\end{picture}
\ca{Bound state}
\label{fig:sboun} } \
\parbox{2.2in}{
\centering
\unitlength 1.00mm
\linethickness{0.4pt}
\begin{picture}(22.00,55.00)(26.00,0)
\put(46.00,3.00){\rule{2.00\unitlength}{52.00\unitlength}}
\color{red}
\multiput(45.67,49.00)(0,1){6}{\circle*{0.25}}
\multiput(45.67,4.00)(0,1){6}{\circle*{0.25}}
\multiput(45.67,11.00)(0,2){19}{\circle*{0.25}}
\color{green}
\put(45.89,48.89){\line(-1,-2){20.06}}
\put(45.89,9.44){\line(-1,2){20.06}}
\color{black}
\end{picture}
\ca{Crossed process}
\label{fig:uboun} }
\end{center}
\end{figure}

Turning now to \m{\xi}-dependent poles and zeroes, we find zeroes at
\be 
u= -\xl +
\frac{(2n+1)\pi}{2\la},
\label{eq:basiczeroes}
\ee 
where \m{n=0,1,2,\ldots}, for \m{P^{+}}, and at the same rapidities but
with \m{\xi \rightarrow -\xi} for \m{P^{-}}. There are also poles in
\m{P^{+}} only at \m{u=\nu_{n}}, with
\be
\fbox{\,$\nu_{n}=\xl - \frac{(2n+1)\pi}{2\la_{\phantom{l}}}$\,}
\ee 
A soliton can only decay into an anti-soliton and a breather, with a
rapidity difference between the two of \m{\hp+\frac{b\pi}{2\la}} for
breather \m{b}. Thus, by lemma~\ref{lemma:2}, all these poles must
correspond to bound states, as shown in \fig{fig:sboun}. For reasons
which will become clear 
in a moment, we
shall depart from the convention of \cite{Skorik} and, rather
than labelling the state corresponding to pole \m{\nu_{n}} as
\m{\beta_{n}}, will label it according to topological charge and \m{n}
as \m{\st{1;n}}. 

\sse{Solitonic excited state reflection factors}
Using the boundary bootstrap
equations given in~\cite{GhoshZam} --- which come from considering
\fig{fig:bootstrap}
 --- solitonic reflection
factors can be calculated for this first set of bound states. In our
case, these equations read
\be
P^{b}_{\st{y}}(u)=
\sum_{c,d}P^{d}_{\st{x}}(u)S^{ab}_{cd}(u-\alpha_{ax}^{y})S^{dc}_{ba}
(u+\alpha_{ax}^{y}),
\ee
where \m{a,b,c}, and \m{d} take the values \m{+} or \m{-} and
\m{\alpha_{ax}^{y}} is the (imaginary) rapidity of the pole at which
particle \m{a} binds to boundary state \m{\st{x}} to give state
\m{\st{y}}. 
The mass of state \m{\st{y}} --- \m{m_{y}} --- is given by
\be
m_{y}=m_{x}+m_{s}\cos \alpha_{ax}^{y}\,.
\ee

\begin{figure}
\begin{center}
\parbox{2.8in}{
\unitlength 1.00mm
\linethickness{0.4pt}
\begin{picture}(48.00,55.00)
\put(46.00,3.00){\rule{2.00\unitlength}{52.00\unitlength}}
\color{green}
\put(30.25,5.53){\line(1,1){15.56}}
\put(45.69,37.08){\line(-1,1){15.56}}
\put(17.89,6.22){\line(4,1){28.00}}
\put(17.89,20.22){\line(4,-1){28.00}}
\color{red}
\multiput(45.78,21.44)(0,1){16}{\circle*{0.25}}
\color{deepred}
\put(28.25,4.53){\makebox(0,0)[cc]{$a$}}
\put(15.89,5.22){\makebox(0,0)[cc]{$b$}}
\put(15.89,21.22){\makebox(0,0)[cc]{$b$}}
\put(28.13,53.64){\makebox(0,0)[cc]{$a$}}
\put(34.75,13.03){\makebox(0,0)[cc]{$c$}}
\put(40.89,9.22){\makebox(0,0)[cc]{$d$}}
\put(42.00,20.00){\makebox(0,0)[cc]{$a$}}
\put(51.00,6.00){\makebox(0,0)[cc]{$\st{x}$}}
\put(51.00,52.00){\makebox(0,0)[cc]{$\st{x}$}}
\put(51.00,29.00){\makebox(0,0)[cc]{$\st{y}$}}
\put(54.00,13.22){\makebox(0,0)[cc]{$P^{d}_{\st{x}}(u)$}}
\color{black}
\end{picture} } \ \LARGE = \normalsize \
\parbox{2.8in}{
\unitlength 1.00mm
\linethickness{0.4pt}
\begin{picture}(48.00,55.00)
\put(46.00,3.00){\rule{2.00\unitlength}{52.00\unitlength}}
\color{green}
\put(30.25,5.53){\line(1,1){15.56}}
\put(45.69,37.08){\line(-1,1){15.56}}
\put(17.89,22.00){\line(4,1){28.00}}
\put(17.89,36.00){\line(4,-1){28.00}}
\color{red}
\multiput(45.78,21.44)(0,1){16}{\circle*{0.25}}
\color{deepred}
\put(28.25,4.53){\makebox(0,0)[cc]{$a$}}
\put(15.89,36.78){\makebox(0,0)[cc]{$b$}}
\put(15.89,21.22){\makebox(0,0)[cc]{$b$}}
\put(28.13,53.64){\makebox(0,0)[cc]{$a$}}
\put(51.00,6.00){\makebox(0,0)[cc]{$\st{x}$}}
\put(51.00,52.00){\makebox(0,0)[cc]{$\st{x}$}}
\put(51.00,23.00){\makebox(0,0)[cc]{$\st{y}$}}
\put(54.00,29.00){\makebox(0,0)[cc]{$P^{b}_{\st{y}}(u)$}}
\color{black}
\end{picture} }
\ca{Boundary bound-state bootstrap}
\label{fig:bootstrap}
\end{center}
\end{figure}

Taking \m{x} to be the ground state \m{\st{0}} and \m{y} to be one
of the set of excited states \m{\st{1;n}}, this gives
\ba
P^{+}_{\st{1;n}}(u)&=&P^{+}_{\st{0}}(u)a(u-\nu_{n})a(u+\nu_{n}) \nn\\
P^{-}_{\st{1;n}}(u)&=&P^{-}_{\st{0}}(u)b(u-\nu_{n})b(u+\nu_{n})+ 
P^{+}_{\st{0}}(u)c(u-\nu_{n})c(u+\nu_{n}).
\label{eq:firstboot}
\ea 
Note that
\m{P^{\pm}_{\st{1;0}}(u)=\overline{P^{\mp}_{\st{0}}(u)}},
where \m{\overline{P^{\pm}(u)}} is \m{P^{\pm}(u)} under the transformation
\m{\xi \rightarrow \pi (\la +1) - \xi}.  The reason for
this is clear
if we look back at \fig{fig:vacua}; this transformation is equivalent
to reflecting the diagram in the horizontal axis, interchanging the
ground and first excited states. 

Perhaps the neatest way to write the new reflection factors is
\be
P^{\pm}_{\st{1;n}}(u) = \overline{P^{\mp}(u)}a^{1}_{n}(u),
\label{eq:newreff}
\ee
where
\be
a^{1}_{n}(u)=\frac{a(u+\nu_{n})a(u-\nu_{n})}{a(u+\nu_{0})a(u-\nu_{0})}.
\ee
The factor \m{a^{1}_{n}(u)} simplifies to
\be
a^{1}_{n}(u)=\prod_{x=1}^{n}\frac{\left( \frac{\xi}{\la
\pi}+\frac{1}{2\la} -\frac{x}{\la} \right)\left( \frac{\xi}{\la \pi} -
\frac{1}{2\la}-\frac{x}{\la}\right)}{ \left( \frac{\xi}{\la
\pi}+\frac{1}{2\la} - \frac{x}{\la} +1\right) \left( \frac{\xi}{\la
\pi} - \frac{1}{2\la} - \frac{x}{\la} +1 \right)}.
\ee

Looking at the pole structure, we find that the functions
\m{\overline{P^{\pm}(u)}} have common simple poles at \m{\nu_{0}} and
\m{\nu_{-N}} where \m{N=1,2,3,\ldots}\,. In addition,
\m{\overline{P^{+}(u)}} has simple poles at 
\m{u=w_{N'}}, where
\be
\fbox{\,$w_{N'}=\pi - \xl -
\frac{\pi(2N'-1)}{2\la_{\phantom{l}}}=\overline{\nu_{N'}}$\,}
\ee
and \m{N'=1,2,3,\ldots}, and simple zeroes at \m{-w_{N''}} for appropriate
values of \m{N''}. Finally, \m{a^{1}_{n}(u)} has simple poles at
\m{\nu_{0}} and \m{\nu_{n}}, and double poles at \m{\nu_{k}},
\m{k=1,2,\ldots,n-1}.

Before proceeding to a more rigorous discussion, we shall now digress
to give an outline of how the bootstrap might 
be expected to work. If one of these poles \emph{does} correspond to 
a new bound state, factorisability leads us to expect that moving the 
soliton and anti-soliton trajectories past each other (so that the 
anti-soliton is incident on the boundary first) should also create the 
same state. The most obvious explanation for this would 
be for the anti-soliton to bind to the boundary first, followed by the 
soliton, to form the state. From above, however, only solitons can bind 
to the ground state, so we must look further.

The \emph{next} most obvious way this could happen is via the soliton 
and anti-soliton forming a breather, either before or after the 
anti-soliton has reflected from the boundary. The poles required to 
allow the first process (of the form \m{\pi +\xl - 
\frac{\pi(2m+1)}{2\la}}) are not present, whereas those necessary 
for the second (of the form \m{w_{m}}) are. Our candidate process 
therefore becomes
\fig{fig:bresoli}, where the soliton and anti-soliton bind to form a 
breather, which then creates the state in one step. It is quite 
difficult to imagine any further alternatives, so let us --- for the 
moment --- take the existence of such a process as a necessary condition 
for a pole to be responsible for the formation of a boundary state.
\begin{figure}
\begin{center}
\parbox{2.5in}{
\unitlength 1.00mm
\linethickness{0.4pt}
\begin{center}
\begin{picture}(55.00,55.00)
\put(46.00,3.00){\rule{2.00\unitlength}{52.00\unitlength}}
\color{green}
\put(25.33,4.60){\line(1,1){20.56}}
\put(25.33,53.29){\line(1,-1){20.56}}
\put(25.33,49.11){\line(2,-1){20.56}}
\put(25.33,9.22){\line(2,1){20.56}}
\color{red}
\multiput(45.78,25.44)(0,1){8}{\circle*{0.25}}
\color{deepred}
\put(23.33,3.50){\makebox(0,0)[cc]{$a$}}
\put(23.33,54.29){\makebox(0,0)[cc]{$a$}}
\put(23.33,8.22){\makebox(0,0)[cc]{$b$}}
\put(23.33,50.11){\makebox(0,0)[cc]{$b$}}
\put(41.33,16.22){\makebox(0,0)[cc]{$c$}}
\put(41.33,43.29){\makebox(0,0)[cc]{$c$}}
\put(41.33,23.60){\makebox(0,0)[cc]{$d$}}
\put(41.33,34.00){\makebox(0,0)[cc]{$d$}}
\put(51.00,6.00){\makebox(0,0)[cc]{$\alpha$}}
\put(51.00,52.00){\makebox(0,0)[cc]{$\alpha$}}
\put(52.00,25.16){\makebox(0,0)[cc]{$g^{\beta}_{\gamma d}$}}
\put(52.00,19.50){\makebox(0,0)[cc]{$g^{\alpha}_{\beta c}$}}
\put(52.00,34.73){\makebox(0,0)[cc]{$x$}}
\put(52.00,39.83){\makebox(0,0)[cc]{$y$}}
\qbezier(46.00,35.73)(45.00,37.73)(43.00,35.73)
\qbezier(46.00,40.83)(45.00,42.83)(43.00,40.83)
\put(50.50,34.73){\vector(-1,0){6}}
\put(50.50,39.83){\vector(-1,0){6}}
\color{black}
\end{picture}
\end{center} } \ \LARGE = \normalsize \
\parbox{2.5in}{
\begin{center}
\unitlength 1.00mm
\linethickness{0.4pt}
\begin{picture}(48.00,55.11)(24.00,0)
\put(46.00,3.00){\rule{2.00\unitlength}{52.00\unitlength}}
\color{red}
\multiput(45.78,25.44)(0,1){8}{\circle*{0.25}}
\color{green}
\put(37.78,3.33){\line(1,1){8.22}}
\put(45.89,11.56){\line(-1,1){5.22}}
\put(25.33,9.22){\line(2,1){15.33}}
\put(37.78,55.11){\line(1,-1){8.22}}
\put(45.89,46.89){\line(-1,-1){5.22}}
\put(25.33,49.22){\line(2,-1){15.33}}
\color{blue}
\multiput(40.67,41.67)(1,-1.66){6}{\circle*{0.25}}
\multiput(40.67,16.78)(1,1.66){6}{\circle*{0.25}}
\color{deepred}
\put(35.78,2.33){\makebox(0,0)[cc]{$b$}}
\put(23.33,8.22){\makebox(0,0)[cc]{$a$}}
\put(23.33,50.22){\makebox(0,0)[cc]{$a$}}
\put(35.78,56.11){\makebox(0,0)[cc]{$b$}}
\put(42.00,23.00){\makebox(0,0)[cc]{$n$}}
\qbezier(46.00,8.55)(45.00,6.55)(43.00,8.55)
\put(50.50,9.55){\vector(-1,0){6}}
\put(52.00,9.55){\makebox(0,0)[cc]{$y$}}
\qbezier(46.00,21.74)(45.00,20.00)(43.50,21.24)
\put(50.50,22.74){\vector(-1,0){6}}
\put(56.00,22.74){\makebox(0,0)[cc]{$x - y$}}
\color{black}
\end{picture}
\end{center} }
\ca{States can be created either by breathers or solitons}
\label{fig:bresoli}
\end{center}
\end{figure}

The consequence of this is that the \m{w_{N}} poles are selected 
as the only possible candidates, and it appears that new bound states can 
only be formed by anti-solitons. Such states hence have charge 0 
(agreeing with the idea that they can also be formed from the ground 
state by the action of a breather). In addition, it is also clear that 
only those \m{w_{N}} such that \m{w_{N}<\nu_{n}} can be considered, as, 
otherwise, the breather version of the process would see the breather 
created heading away from the boundary, rather than towards it.

Designating such a new state as \m{\st{0;n,N}} and bootstrapping on it
leads to
\ba
P^{-}_{\st{0;n,N}}(u)&=&P^{-}_{\st{1;n}}(u)a(u-w_{N})a(u+w_{N}) \nn\\
P^{+}_{\st{0;n,N}}(u)&=&P^{+}_{\st{1;n}}(u)b(u-w_{N})b(u+w_{N})+ 
P^{-}_{\st{1;n}}(u)c(u-w_{N})c(u+w_{N}).
\ea 
Substituting in \re{eq:newreff} and taking advantage of the fact that 
\m{w_{N}=\overline{\nu_{N}}} (so \m{a(u\pm w_{N})=\overline{a(u\pm 
\nu_{N})}}), this becomes
\ba
P^{-}_{\st{0;n,N}}(u)&=&a_{n}^{1}(u)\overline{P^{+}_{\st{0}}(u)}
a(u-\overline{\nu_{N}})a(u+\overline{\nu_{N}}) \nn \\
P^{+}_{\st{0;n,N}}(u)&=&a_{n}^{1}(\overline{P^{-}_{\st{0}}(u)}
b(u-\overline{\nu_{N}})b(u+\overline{\nu_{N}})+ 
\overline{P^{+}_{\st{0}}(u)}c(u-\overline{\nu_{N}})c(u+\overline{\nu_{N}})),
\ea
which (apart from an extra factor of \m{a_{n}^{1}(u)}) is just the first 
bootstrap \re{eq:firstboot} under the transformation \m{\xi \rightarrow 
\pi(\la+1)-\xi} and with solitons and anti-solitons interchanged on the 
lhs. Thus, the pole structure follows naturally from the above. This can 
also be written as
\be
P^{\pm}_{\st{0;n,N}}(u) = P^{\pm}_{\st{0}}(u)a_{n}^{1}(u)
\overline{a_{N}^{1}(u)}.
\ee

Repeating the factorisation argument shows that now we should focus on 
\m{\nu_{n'}} poles such that \m{\nu_{n'}<w_{N}}. These are present now 
in the solitonic factor, though (due to the extra factor of 
\m{a_{n}^{1}(u)}) only for \m{n'>n}. However, since any such state obeys 
\m{\nu_{n}>w_{N}>\nu_{n'}} in any case, this restriction is not relevant.
The resultant state must now have charge 0.

A pattern is emerging, and it is not hard to see how the process 
would continue. Starting from the ground state, and taking the broadest 
guess (given our assumptions) for the spectrum, states can be formed by 
alternating solitons and anti-solitons, the solitons having rapidity 
\m{\nu_{n_{i}}} and the anti-solitons having rapidity \m{w_{N_{j}}} (for 
some sets \m{\underline{n}} and \m{\underline{N}}). 
An schematic pole structure is shown in \fig{fig:poles}, in terms of which 
the criterion
for a state to be in the spectrum should be that we begin with one of the
\m{\nu_{n}} and then, as we move along the index list, move down
the diagram, switching from side to side as we go. If we finish on a
\m{\nu_{m}} (indicating that the most recent particle to bind was a
soliton) the state has charge 1 while, if we finish on a \m{w_{m}}
(meaning an anti-soliton) the state has charge 0.
\begin{figure}
\begin{center}
\unitlength=1.00mm
\begin{picture}(82.35,53.75)(15.00,0.00)
\color{deepred}
\put(40.00,53.75){\line(0,-1){53.75}}
\put(20.00,53.75){\line(0,-1){53.75}}
\put(19.00,53.75){\line(1,0){2}}
\put(19.00,0.00){\line(1,0){2}}
\put(16.00,52.75){$\hp$}
\put(16.00,-1.00){$0$}
\put(16.00,25.87){$u$}
\multiput(30.00,47.00)(1,0){21}{\circle*{0.25}}
\put(52.00,47.00){\makebox(0,0)[cc]{$\xl$}}
\put(58.00,32.00){\makebox(0,0)[cl]{$\nu_{n}=\xl-\frac{\pi(2n+1)}{2\la}$}}
\put(58.00,23.00){\makebox(0,0)[cl]{$w_{m}=\pi-\xl-\frac{\pi(2m-1)}{2\la}$}}
\color{blue}
\multiput(40.00,42.00)(0,-10){5}{\line(1,0){5}}
\multiput(40.00,42.00)(0,-10){5}{\circle*{2}}
\put(46.00,41.00){$\nu_{0}$}
\put(46.00,31.00){$\nu_{1}$}
\put(46.00,21.00){$\nu_{2}$}
\put(46.00,11.00){$\nu_{3}$}
\put(46.00,1.00){$\nu_{4}$}
\color{red}
\multiput(40.00,49.00)(0,-10){5}{\line(-1,0){5}}
\multiput(40.00,49.00)(0,-10){5}{\circle*{2}}
\put(30.00,48.00){$w_{2}$}
\put(30.00,38.00){$w_{3}$}
\put(30.00,28.00){$w_{4}$}
\put(30.00,18.00){$w_{5}$}
\put(30.00,8.00){$w_{6}$}
\color{black}
\end{picture}
\ca{Location of poles. \mdseries (Note that, in this case, $w_{2}$ can
never participate in bound state formation as it is above $\nu_{0}$.)}
\label{fig:poles} 
\end{center}
\end{figure}     

Annotating such a state by its topological charge, \m{c}, and the sets 
\m{\underline{n}} and \m{\underline{N}} as 
\m{\st{c;n_{1},N_{1},n_{2},N_{2},\ldots}} (noting 
\m{\nu_{n_{1}}>w_{N_{1}} >\nu_{n_{2}} >w_{N_{2}}>\ldots}), the solitonic 
reflection factors should be
\be
P^{\pm}_{\st{c;n_{1},N_{1},\ldots}}(u)=P^{\pm}_{(c)}(u)a^{1}_{n_{1}}(u)
\overline{a^{1}_{N_{1}}(u)}\ldots\,, 
\ee
with \m{P^{\pm}_{0}(u)=P^{\pm}_{\st{0}}(u)} and \m{P^{\pm}_{1}(u)=
\overline{P^{\pm}_{\st{0}}(u)}}. From now on, however, it will be more
convenient to consider a single index list, and denote 
\m{\overline{a^{1}_{m}(u)}} as \m{a^{0}_{m}(u)}, giving
\be
P^{\pm}_{\st{c;n_{1},n_{2},\ldots,n_{k}}}(u)=P^{\pm}_{(c)}(u)a^{1}_{n_{1}}(u)
a^{0}_{n_{2}}(u)a^{1}_{n_{3}}(u)\ldots a^{c}_{n_{k}}(u), 
\ee 
where \m{c} is 1 if \m{k} is odd, and 0 if \m{k} is even. We will call 
this a level \m{k} boundary bound state.
If we choose the ground state mass to be 
\m{m_s\sin^{2}\left(\frac{\xi-\hp}{2\la}\right)}, the mass of this
state is
\ba
m_{n_{1},n_{2},\ldots}&=&
m_s\sin^{2}\left(\frac{\xi-\hp}{2\la}\right)+
\sum_{i \mathrm{\ odd}} m_s\cos(\nu\phup_{n_i})
+ \sum_{j
\mathrm{\ even}} m_s\cos(w\phup_{n_j})\\[4pt]
&=&
m_s\sin^{2}\left(\frac{\xi-\hp}{2\la}\right)+
\sum_{i \mathrm{\ odd}} m_s\cos \left(
\xl - \frac{(2n_{i}+1)\pi}{2\la}\right) - \sum_{j
\mathrm{\ even}} m_s\cos \left( \xl+
\frac{(2n_{j}-1)\pi}{2\la} \right)\,.\nn
\label{eq:mass}
\ea
This choice is convenient in that, as \m{\xi} passes \m{\pi/\beta},
the masses of the ground and first excited states interchange, in line
with the idea that the states themselves swap at this point.
An important point to note is that, in deriving all this, we have simply 
been considering the soliton sector. However, we will see that allowing
breather processes as well does not give rise to any further states, 
merely additional ways to jump between states. The Dirichlet boundary 
condition is also special in that either the soliton or the anti-soliton 
can couple to a given boundary, but not both, as might be generically 
expected.

Although we have built up the states by applying the solitons and
anti-solitons in this alternating fashion, precisely how this happens in a
given situation
will of course depend on the impact
parameters of the incoming particles. 
In \fig{fig:bresoli} we already gave an example of the
complicated way in which a process may be rearranged as these impact
parameters vary, and the particular choices that we have adopted are
mainly motivated by a desire to assemble the full spectrum in the simplest
possible way.

\sse{Breather ground state reflection factors}
We now return to the pole analysis, and examine the breather ground state
reflection factors (\ref{gsrfl}).
Again, the factor \m{R^{n}_{0}} is boundary-independent, and so all 
its poles should have an explanation in terms of the bulk. There are 
(physical strip) poles at \m{\hp}, \m{\frac{l\pi}{2\la}}, 
\m{\hp-\frac{n\pi}{2\la}}, and double poles at \m{\hp-\frac{l\pi}{2\la}}, 
with \m{l=1,2,\ldots,n-1}. There are no zeroes.
The pole at \m{\hp} is simply due to the breather coupling
perpendicularly to the boundary, while the poles at 
\m{\frac{l\pi}{2\la}} are explained by \fig{fig:bretri}. Next, the pole at 
\m{\hp-\frac{n\pi}{2\la}} comes from a breather version of 
\fig{fig:bbre}, \m{B_{2n}} being formed. Finally, the double poles at 
\m{\hp - \frac{l\pi}{2\la}} are due to \fig{fig:bredoubletri}.

\begin{figure}
\begin{center}
\parbox{3.4in}{ 
\centering
\unitlength 1.00mm
\linethickness{0.4pt}
\begin{picture}(38.00,55.00)(10.00,0)
\put(46.00,3.00){\rule{2.00\unitlength}{52.00\unitlength}}
\color{blue}
\multiput(46.00,29.00)(-1,-0.5){21}{\circle*{0.25}}
\multiput(46.00,29.00)(-1,0.5){21}{\circle*{0.25}}
\multiput(26.00,19.00)(0,1){21}{\circle*{0.25}}
\multiput(26.00,19.00)(-1,-1){16}{\circle*{0.25}}
\multiput(26.00,39.00)(-1,1){16}{\circle*{0.25}}
\color{deepred}
\put(20.00,29.00){\makebox(0,0)[cc]{$n-l$}}
\put(36.00,21.00){\makebox(0,0)[cc]{$l$}}
\put(36.00,37.00){\makebox(0,0)[cc]{$l$}}
\put(18.50,8.50){\makebox(0,0)[cc]{$n$}}
\put(18.50,49.50){\makebox(0,0)[cc]{$n$}}
\qbezier(26.00,23.00)(29.00,24.00)(30.00,21.00)
\put(31.00,17.00){\vector(-1,1){4}}
\put(34.00,14.50){\makebox(0,0)[cc]{$\frac{n\pi}{2\la}$}}
\color{black}
\end{picture}
\ca{Breather triangle process}
\label{fig:bretri} } \
\parbox{3.4in}{ 
\centering
\unitlength 1.00mm
\linethickness{0.4pt}
\begin{picture}(37.00,55.00)(11.00,0)
\put(46.00,3.00){\rule{2.00\unitlength}{52.00\unitlength}}
\color{blue}
\multiput(31.00,44.00)(1,0){16}{\circle*{0.25}}
\multiput(31.00,14.00)(1,0){16}{\circle*{0.25}}
\multiput(31.00,44.00)(1,-1){16}{\circle*{0.25}}
\multiput(31.00,14.00)(1,1){16}{\circle*{0.25}}
\multiput(31.00,14.00)(-1,-0.5){21}{\circle*{0.25}}
\multiput(31.00,44.00)(-1,0.5){21}{\circle*{0.25}}
\color{deepred}
\put(38.5,47.00){\makebox(0,0)[cc]{$n-l$}}
\put(38.5,11.00){\makebox(0,0)[cc]{$n-l$}}
\put(38.5,33.5){\makebox(0,0)[cc]{$l$}}
\put(38.5,24.5){\makebox(0,0)[cc]{$l$}}
\put(21.00,52.00){\makebox(0,0)[cc]{$n$}}
\put(21.00,6.00){\makebox(0,0)[cc]{$n$}}
\qbezier(34.00,17.00)(36.50,16.50)(35.00,14.00)
\put(29.00,19.00){\vector(1,-1){4}}
\put(26.00,22.00){\makebox(0,0)[cc]{$\frac{n\pi}{2\la}$}}
\color{black}
\end{picture}
\ca{Breather double triangle process}
\label{fig:bredoubletri} }
\end{center}
\end{figure}

Moving on to the boundary-dependent part, there are 
poles at
\be
u=\xl - \hp \pm \frac{l\pi}{2\la},
\ee
and zeroes at
\ba
u&=&-\xl +\hp \pm \frac{l\pi}{2\la} \nonumber\\
u&=&\xl +\hp \pm \frac{l\pi}{2\la}\,,
\ea
where, for a breather \m{n}, \m{l=n-1, n-3,\ldots, l \geq 0}.

The set of poles can be re-written by noting
that, for breather \m{m}, there is a simple pole of the form
\m{\hf(\nu_{n}-w_{N})} for all \m{n,N \geq 0} and \m{n,N \in Z}
such that \m{m=n+N}. This ties in with the discussion in the previous 
section, since these are the rapidities predicted for the single-step 
process which is equivalent to a soliton binding at an angle of 
\m{\nu_{n}} followed by an anti-soliton at \m{w_{N}}.

\sse{Breather excited state reflection factors}
Following the discussion of the solitonic excited state
reflection factors, we can introduce corresponding breather reflection 
factors:
\be
R^{m}_{\st{c;n_{1},n_{2},\ldots,n_{k}}}(u)=R^{m}_{(c)}(u)a^{1;m}_{n_{1}}(u)
a^{0;m}_{n_{2}}(u)a^{1;m}_{n_{3}}(u)\ldots a^{c;m}_{n_{k}}(u), 
\label{eq:breexc}
\ee 
where
\m{R^{m}_{0}(u)=R^{m}_{\st{0}}(u)} and \m{R^{m}_{1}(u)= \overline{R^{m}_{
\st{0}}(u)}}. We have also defined 
\be
a^{c;m}_{n}(u)=a_{n}^{c}\left(u+\frac{u_{m}}{2}\right)a_{n}^{c}\left(u
-\frac{u_{m}}{2}\right),
\label{eq:simpleacn}
\ee
or
\be
a^{1;m}_{n}(u)=\prod_{x=1}^{m}\frac{\left( \frac{\xi}{\la
\pi}+\frac{1-2x-n}{2\la}+\hf\right)\left( \frac{\xi}{\la
\pi}-\frac{1+2x+n}{2\la}+\hf\right)}{ \left( \frac{\xi}{\la
\pi}+\frac{1-2x-n}{2\la}-\hf\right) \left( \frac{\xi}{\la \pi} -
\frac{1+2x+n}{2\la} - \hf\right)} \frac{\left(\frac{\xi}{\la \pi}+
\frac{1-2x+n}{2\la} -\hf\right)\left( \frac{\xi}{\la \pi}
-\frac{1+2x-n}{2\la} -\hf \right)}{\left( \frac{\xi}{\la \pi}+
\frac{1-2x+n}{2\la} +\hf \right)\left( \frac{\xi}{\la \pi} - \frac{1+
2x-n}{2\la} + \hf \right)},
\label{eq:complexacn}
\ee
with \m{a^{0;m}_{n}(u)=\overline{a^{1;m}_{n}(u)}}.

For \m{\overline{R^{m}_{\st{0}}(u)}}, there are poles at
\ba
u&=&\hp-\xl+\frac{\pi}{\la}\pm\frac{l\pi}{2\la} \nonumber \\
u&=&\hp+\xl-\frac{(l+2)\pi}{2\la}\,,
\ea
and zeroes at
\be
u=\xl-\hp+\frac{(l-2)\pi}{2\la}\,.
\ee

For the other factors, \m{a^{1;m}_{n}(u)} has
physical strip poles/zeroes at
\be
\begin{array}{crl}
u = -\xl + \hp + \frac{p\pi}{2\la} & \mathrm{poles:\
} & p = 2n-m+2x \pm 1 \\
 & \mathrm{zeroes:\ } & p = - m +2x \pm 1 \\
u = \xl - \hp + \frac{p\pi}{2\la} & \mathrm{poles:\
} & p = m-2x\pm 1 \\
 & \mathrm{zeroes:\ } & p = -2n+m-2x\pm 1 \\
u = \xl + \hp + \frac{p\pi}{2\la} & \mathrm{poles:\
} & p = -2n+m-2x\pm 1 \\
 & \mathrm{zeroes:\ } & p = m-2x\pm 1
\end{array}
\ee
while \m{a^{0;m}_{n}(u)} has them at
\be
\begin{array}{crl}
u = -\xl + \frac{3\pi}{2} + \frac{p\pi}{2\la} & \mathrm{poles:\
} & p = -2N-m+2x\pm 1 \\
 & \mathrm{zeroes:\ } & - \\
u = -\xl + \hp + \frac{p\pi}{2\la} & \mathrm{poles:\
} & p = -m+2x \pm 1 \\
 & \mathrm{zeroes:\ } & p = -2N - m +2x \pm 1 \\
u = \xl - \hp + \frac{p\pi}{2\la} & \mathrm{poles:\
} & p = 2N + m-2x\pm 1 \\
 & \mathrm{zeroes:\ } & p = m-2x\pm 1 \\
u = \xl + \hp + \frac{p\pi}{2\la} & \mathrm{poles:\
} & p = m-2x\pm 1 \\
 & \mathrm{zeroes:\ } & p = 2N + m-2x\pm 1
\end{array}
\ee
These poles will be further discussed 
in \sect{sec:general} below.

\se{An example}
\label{sec:example}
To get an idea of the full picture, and which processes are 
responsible for the remaining poles,
we will now look at one particular example in more detail. If we select
\m{\xi=1.6\pi} and \m{\la=2.5}, then we have the first two breathers in
the spectrum, with the solitonic poles taking the form \m{\nu_{n} =
\frac{\pi(2.2-2n)}{5}} and \m{w_{N}=\frac{\pi(2.8-2N)}{5}}. 
Thus, for this case, only the poles at \m{\nu_{0}, \nu_{1}} and
\m{w_{1}} are on the physical strip, and so \fig{fig:poles} is
simplified to \fig{fig:expoles}. This is
the simplest case which requires a broader spectrum than that
postulated in \cite{Skorik}. First, let us turn to the soliton
sector.
\begin{figure}
\begin{center}
\unitlength=1.00mm
\begin{picture}(32.35,53.75)(15.00,0.00)
\color{deepred}
\put(40.00,53.75){\line(0,-1){53.75}}
\put(20.00,53.75){\line(0,-1){53.75}}
\put(19.00,53.75){\line(1,0){2}}
\put(19.00,0.00){\line(1,0){2}}
\put(16.00,52.75){$\hp$}
\put(16.00,-1.00){$0$}
\put(16.00,25.87){$u$}
\color{blue}
\multiput(40.00,47.30)(0,-43){2}{\line(1,0){5}}
\multiput(40.00,47.30)(0,-43){2}{\circle*{2}}
\put(46.00,47.30){$\nu_{0}$}
\put(46.00,4.3){$\nu_{1}$}
\color{red}
\put(40.00,17.20){\line(-1,0){5}}
\put(40.00,17.20){\circle*{2}}
\put(30.00,17.20){$w_{1}$}
\color{black}
\end{picture}
\ca{Location of poles in the example}
\label{fig:expoles}
\end{center}
\end{figure}

\sse{Boundary ground state --- soliton sector}
As argued above, the soliton can bind to the boundary at all
rapidities \m{\nu_{n}} which are in the physical strip, here just
comprising \m{\nu_{0}} and \m{\nu_{1}}.  This introduces the states
\m{\st{1;0}} and \m{\st{1;1}}.

\sse{Boundary ground state --- breather sector}
The only breather poles are at \m{\xl-\hp+\frac{(m-1)\pi}{2\la}} for
breather \m{m}. In addition, breather \m{B_{2}} has a zero at
\m{-\xl+\hp+\frac{\pi}{2\la}}.

By lemma~\ref{lemma:1}, the pole for \m{B_{1}} must correspond to a
new bound state, the rapidity being less than
\m{\frac{\pi}{2\la}}. From \fig{fig:bresoli},
it is clear that \m{B_{1}} creates the
state which was labelled \m{\st{\delta_{0,1}}} in \cite{Skorik},
and which we have called \m{\st{0;0,1}}.

The pole for the second breather can be explained by
\fig{fig:brfake}, with the state \m{\st{1;0}} being formed. The
anti-soliton is reflected from the boundary at a rapidity of
\m{\xl-\pi+\frac{3\pi}{2\la}} --- a zero of the \m{\st{1;0}} reflection
factor --- reducing the diagram to first order through the boundary
Coleman-Thun mechanism.

\sse{First level excited states --- soliton sector}
{}From before, \m{P^{+}_{\st{1;0}}} just has a simple pole at
\m{\nu_{0}}, which
can be explained by the crossed process in \fig{fig:uboun}, reducing
the boundary to the ground state.
For \m{P^{+}_{\st{1;1}}}, the pole
at \m{\nu_{1}} can be explained this way while, for the double pole at
\m{\nu_{0}}, \fig{fig:fb4} is required, the first breather being
formed while the boundary is reduced to the vacuum state.

For \m{P^{-}_{\st{1;n}}(u)}, we have the additional job of explaining
simple poles at \m{w_{N}}, for all \m{N} such that this pole is in the
physical strip. Here, this is only \m{w_{1}}. For \m{\st{1;0}}, this is
appropriate for the formation of \m{\st{0;0,1}} which, from the previous
section, must be present. For \m{\st{1;1}}, however, \fig{fig:fb2} is
invoked, the second breather being created, and the boundary reduced
to the vacuum state. The breather is incident on the boundary at an
angle of \m{\hf(w_{1}-\nu_{1})=\pi - \xl - \frac{\pi}{2\la}} which, looking
at the above breather reflection factors, is a zero, ensuring the
diagram is of the correct order.

\sse{First level excited states --- breather sector}
The pole structure of
\m{{R_{\st{1;0}}^{m}}} 
can be found from \m{\overline{R_{\st{0}}^{m}}}, and is
\be
\begin{array}{rrl}
B_{1}: & \mathrm{pole\ at\ } & \hp-\xl+\frac{\pi}{\la} \\
B_{2}: & \mathrm{poles\ at\ } & \hp - \xl+\frac{3\pi}{2\la},
\hp-\xl+\frac{\pi}{2\la}
\end{array}
\ee

By
lemma~\ref{lemma:1}, the second pole for \m{B_{2}} must correspond to
a new bound state; by the previous arguments, this is
\m{\st{1;0,1,1}}. 
This state is not in the spectrum given in~\cite{Skorik}, 
but lemma~\ref{lemma:1} shows
that there is no way to avoid its introduction. Considerations such as
this will open the door to a much wider spectrum 
in the general case.

The \m{B_{1}} pole is suitable for the creation of \m{\st{1;1}}. The
first pole for \m{B_{2}} can be explained by \fig{fig:ob6}, with the
boundary being reduced to the ground state by emission of a soliton.

For \m{R_{\st{1;1}}^{m}}, the above poles are supplemented by
additional poles from
\m{a_{1}^{1;m}(u)} to give the poles shown in table \ref{tab:bre1poles}.
\begin{table}[h]
\[
\begin{array}{c|c|c|c} \thline
 & -\xl+\hp+\frac{p\pi}{2\la} &
\xl-\hp+\frac{p\pi}{2\la} &
\xl+\hp+\frac{p\pi}{2\la} \\
\thline B_{1} &2 &0 &- \\
B_{2} &3^{2} &1 &-5 \\
\thline
\end{array}
\]
\vspace{-0.25in}
\ca{Breather pole structure for $\mathbf{\st{1;1}}$. \mdseries Entries
are the values of
\m{p} for which there is a pole in the location given in the column
heading. The power of the entry
gives the order of the pole, so e.g. \m{3^{2}} indicates a double pole
when \m{p=3}. There are no physical strip zeroes for either breather.
}
\label{tab:bre1poles}
\end{table}

The pole at \m{\xl+\hp-\frac{5\pi}{2\la}}
can be explained by \fig{fig:ob4}, with the boundary being reduced to
the ground state by emission of a soliton. The pole at \m{\xl-\hp} for
\m{B_{1}} can be allocated to the creation of \m{\st{1;0,1,1}}, while
the pole at \m{\xl-\hp+\frac{\pi}{2\la}} for \m{B_{2}}
is due to \fig{fig:ob5}, where the boundary emits \m{B_{1}}, being reduced to
\m{\st{1;0}}. The pole at \m{-\xl+\hp+\frac{2\pi}{2\la}} for \m{B_{1}}
is responsible for
this reduction to \m{\st{1;0}}, while the double pole for \m{B_{2}}
comes from an all-breather version of \fig{fig:ob3}, the boundary
being reduced in the same way.

\sse{Second level excited states --- soliton sector}

For \m{P^{-}_{\st{0;0,1}}(u)}, the only poles are simple,
at \m{\nu_{0}} and \m{w_{1}}. The pole at \m{w_{1}} can be explained
by \fig{fig:uboun} while, for \m{\nu_{0}}, we need \fig{fig:fb3}. The
second breather is emitted by the boundary, reducing it to the ground
state, while a soliton is incident on the boundary at a
rapidity \m{w_{1}}. For the ground state, this is neither a pole
nor a zero, but the diagram contains a solitonic loop which can either
be drawn to leave a soliton or an anti-soliton incident on the
boundary. Adding the contributions of these two diagrams gives an
additional zero.

For \m{P^{+}_{\st{0;0,1}}(u)}, we have additional poles at all \m{\nu},
i.e. a simple pole at \m{\nu_{1}}, with
\m{\nu_{0}} becoming a double pole.
By lemma~\ref{lemma:1}, \m{\nu_{1}} must correspond to
the creation of a new bound state, namely \m{\st{1;0,1,1}}, while, for
\m{\nu_{0}}, \fig{fig:fb2} should be considered. Again, the second
breather is created, the boundary is reduced to the ground state, and
the breather is incident on the boundary at a rapidity of
\m{\hf(\nu_{0}-w_{1})=\xi/\la-\pi/2} --- a zero of the
reflection factor.

\sse{Second level excited states --- breather sector}

For \m{\st{0;0,1}}, we have the pole structure given in table
\ref{tab:bre2poles}.
\begin{table}[h]
\[
\begin{array}{c|c|c|c} \thline
 & -\xl+\frac{3\pi}{2} +\frac{p\pi}{2\la} &
-\xl+\hp+\frac{p\pi}{2\la} &
\xl-\hp+\frac{p\pi}{2\la} \\
\thline B_{1} &-2 &2 &0,2 \\
B_{2} &-3 &3 &1^{2} \\
\thline
\end{array}
\]
\vspace{-0.25in}
\ca{Breather pole structure for $\mathbf{\st{0;0,1}}$.}
\label{tab:bre2poles}
\end{table}

The poles at \m{-\xl+\frac{3\pi}{2}+\frac{p\pi}{2\la}} are due to
\fig{fig:ob4}, while the poles in the second column are due to
\fig{fig:ob5}. For all these, the boundary is reduced to \m{\st{1;0}}.
The pole at \m{\xl-\hp+\frac{(m-1)\pi}{2\la}} for \m{B_{m}} (\m{m=2})
is due to \fig{fig:ob7}, while for \m{m=1} it is due to a breather
version of \fig{fig:uboun}. The pole at \m{\xl-\hp+\frac{2\pi}{2\la}}
for \m{B_{1}} is due to \fig{fig:ob6}.

\pagebreak

\sse{Third level excited states --- soliton sector}
The only third level excited state is \m{\st{1;0,1,1}}. For
\m{P^{+}_{\st{1;0,1,1}}}, there are simple poles at
\m{w_{1},\nu_{0}} and \m{\nu_{1}}. Again, the pole at \m{w_{1}} comes from 
the crossed process \fig{fig:uboun}. For \m{\nu_{1}}, \fig{fig:uboun}
suffices while, for \m{\nu_{0}}, \fig{fig:fb4} is required, the
boundary being reduced to \m{\st{0;0,1}} while the first breather is
incident on the boundary at \m{\hp-\xl+\frac{\pi}{\la}}, another zero.

\sse{Third level excited states --- breather sector}
Here, the only possible boundary state is \m{\st{1;0,1,1}} and we find
the poles given in table \ref{tab:bre3poles}.
\begin{table}[h]
\[
\begin{array}{c|c|c|c|c} \thline
 & -\xl+\frac{3\pi}{2} +\frac{p\pi}{2\la} &
-\xl+\hp+\frac{p\pi}{2\la} &
\xl-\hp+\frac{p\pi}{2\la} & \xl+\hp+\frac{p\pi}{2\la} \\
\thline B_{1} &-2 &2^{2},4 &0,2 &- \\
B_{2} &-3 &1,3^{3} &1^{2} &-5 \\
\thline
\end{array}
\]
\vspace{-0.25in}
\ca{Breather pole structure for $\mathbf{\st{1;0,1,1}}$.}
\label{tab:bre3poles}
\end{table}

Comparing this with the structure given above for
\m{\st{1;1}}, it can easily be seen that, whenever the two both have a pole
at the same rapidity, essentially the same explanation can be used. For the
remaining poles, \m{-\xl+\frac{3\pi}{2}+\frac{p\pi}{2\la}} can be
explained by \fig{fig:ob5}, the boundary being reduced to \m{\st{1;0}},
while that at \m{-\xl+\hp+\frac{\pi}{2\la}} for \m{B_{2}} is due to
\fig{fig:uboun}, reducing the boundary to \m{\st{1;0}}, and that at
\m{\xl-\hp+\frac{2\pi}{2\la}} for \m{B_{1}} is due to an all-breather
version of \fig{fig:ob6}, again reducing the boundary to \m{\st{1;0}}.

\sse{Summary}
The above has shown that, by introducing only the states which are
required by lemmas 1 and 2, the complete pole structure can be explained.
Below, we shall find that this is a general feature. In addition, the
spectrum of states is broader than that introduced in~\cite{Skorik}
(containing, in addition to their states, \m{\st{1;0,1,1}}).
It should be noted that the mass of this extra state corresponds
to \m{m_{1,1}} of \cite{Skorik}, the mass of a boundary Bethe ansatz
(1,1)-string whose apparent
absence from the bootstrap spectrum was described in that paper as
``confusing''. 
This does at least show that the Bethe ansatz
results of \cite{Skorik} are not incompatible with
the bootstrap. However, in more general cases it turns out that the
bootstrap predicts yet further states, beyond those identified in the 
boundary Bethe ansatz calculations of \cite{Skorik}, so a full reconciliation
of the Bethe ansatz and bootstrap approaches remains an open problem.

\se{The general case}
\label{sec:general}
{}From the above, we might be tempted to guess that the boundary state
\m{\st{c;n_{1},n_{2},n_{3},\ldots,n_{m}}} exists iff \m{c} is \m{0} or
\m{1} and \m{n_{1},n_{2},n_{3},\ldots} are chosen such that
\m{\pi/2>\nu_{n_{1}}>w_{n_{2}}>\nu_{n_{3}}>\ldots>0}. This turns out to be 
correct, and will be proved in two stages. Firstly, we need to
show that all these states must be present, before going on to show
that, given this, all other poles can be explained without invoking
further boundary states.

\sse{The minimal spectrum}
The argument proceeds as follows: starting with the knowledge that
the vacuum state \m{\st{0}} and all appropriate states
\m{\st{1;n_{1}}} are in the spectrum, we use breather poles to
construct all the other postulated states.

These poles are of the form \m{\hf(w_{N}-\nu_{n})} for breather
\m{n+N} incident on a charge 0 state (or \m{\hf(\nu_{n}-w_{N})}
for a charge 1 state). If \m{\nu_{n}-w_{N}<\pl}, lemma~\ref{lemma:1}
shows that they must correspond either to the formation of a new
state, or the crossed process. From \fig{fig:bresoli}, this
corresponds either to adding indices \m{n} and \m{N} if they are
absent or --- if they are already present --- removing them. (Note that
any other option would give rise to a state with a mass outside the
scheme given by \re{eq:mass}, and therefore outside our postulated
spectrum.) The condition \m{\nu_{n}-w_{N}<\pl} is always satisfied
if \m{\nu_{n}>w_{N}} and \m{\nu_n} and \m{w_N}
are as close together as possible,
i.e. if \m{\st{0;n,N}} exists, but \m{\st{0;n,N-1}} does not.

The only subtlety in this argument arises when considering the topmost
breather. If \m{n+N=n_{max}}, lemma~\ref{lemma:1} on its own is not
strong enough to require the presence of the state we need, and we
must invoke the stronger version introduced at
the end of section~\ref{sec:colethun}. This makes use the idea that there
must be a corresponding two-stage solitonic route to the same state, i.e. a
soliton with rapidity $\nu_{n}$ followed by an anti-soliton with
rapidity $w_{N}$. Considering these two processes instead, the
stronger lemma shows that both form bound states, as $\nu_{n}$ and
$w_{N}$ must be
the lowest poles of their type --- and so have rapidity less than
$\frac{\pi}{\la}$ --- for $n+N$ to equal $n_{max}$. This shows that
the state exists, and hence that the breather pole is due to its formation.

Since the arguments for the two sectors are analogous, let us focus on
the charge 0 sector here. The challenge is to create any state
\m{\st{0;\underline{x}}} --- for some set of indices
\m{\underline{x}=(n_{1},n_{2},\ldots,n_{2k})} --- from the ground state
using just these poles. As a first step, consider creating
\m{\st{0;n_{1},n_{2}}}. If \m{\nu_{n_1}} and \m{w_{n_2}} 
are as close together as possible, we
simply make use of the pole at
\m{\hf(w_{n_{2}}-\nu_{n_{1}})}. Otherwise, introduce the set
\m{m_{1},m_{2},\ldots,m_{t}} such that
\m{\nu_{n_{1}}>w_{m_{1}}>\nu_{m_{2}}>w_{m_{3}}>\ldots
>\nu_{m_{t}}>w_{n_{2}}}, with each successive rapidity as close to the
previous one as possible.  Now, we can successively create
\m{\st{0;\underline{x},n_{1},m_{1}}}, then \m{\st{0;x,n_{1},m_{1},
m_{2}, m_{3}}}
and so on, up to \m{\st{0;\underline{x},n_{1},m_{1},m_{2},m_{3}, 
\ldots,m_{t},n_{2}}}.

By now invoking the crossed process, a
suitable breather can be used to removed the indices \m{m_{1},m_{2}},
followed by \m{m_{3},m_{4}} and so on, until all the \m{m} indices
have been removed to leave \m{\st{0;\underline{x},n_{1},n_{2}}}.

Repeating this procedure allows \m{\st{0;n_{1},n_{2},n_{3},n_{4}}} to
be created, and hence \m{\st{0;\underline{x}}}. Note that this
allows any state in our allowed spectrum to be created, but no others,
as the condition \m{\nu_{n_{1}}>w_{n_{2}}>\ldots} is imposed by the
existence of the necessary breather poles. Charge 1 states can be
created analogously by starting from a suitable state \m{\st{1;n_{1}}}.

One remaining point is to check that all the necessary breather
poles do indeed exist. However, starting from \re{eq:breexc}, they occur in
the \m{R^{n}_{(c)}(u)} factor, and it is straightforward to check that
they are never modified by the other \m{a} factors.

\sse{Reflection factors for the minimal spectrum}
\label{sec:refmin}

The boundary state can be changed by the solitonic
processes given in table \ref{tab:solproc}.
\begin{table}[h]
\[
\begin{array}{c|c|c|c} \thline
\mathbf{Initial\ state} & \mathbf{Particle} & \mathbf{Rapidity} &
\mathbf{Final\ state} \\
\thline \st{0;n_{1},\ldots,n_{2k}} & \mathrm{Soliton} &
\nu_{n} & \st{1;n_{1},\ldots,n_{2k},n} \\
\st{1;n_{1},\ldots,n_{2k-1}} &
\mathrm{Anti-soliton} & w_{N} & \st{0;n_{1},\ldots, n_{2k-1},N} \\
\thline
\end{array}
\]
\vspace{-0.25in}
\ca{Solitonic processes which change the boundary state.}
\label{tab:solproc}
\end{table}

The breather sector is more complex, as indices can be added or
removed from any point in the list, and not just at the end, as for
solitons. In addition, processes exist which simply adjust the value
of one of the indices, rather than increasing the number of
indices. For breather \m{m}, these are given in table \ref{tab:breproc}.
\begin{table}[h]
\[
\begin{array}{c|c|c} \thline
\mathbf{Initial\ state} & \mathbf{Rapidity} & \mathbf{Final\ state} \\
\thline \st{0/1;n_{1},\ldots,n_{2x},n_{2x+1},\ldots} & \hf(\nu_{n}-w_{N}),
n+N=m & \st{0/1;n_{1},\ldots,n_{2x},n,N,n_{2x+1},\ldots} \\
\st{0/1;n_{1},\ldots,n_{2x-1},n_{2x},\ldots} & \hf(w_{N}-\nu_{n}),
n+N=m & \st{0/1;n_{1},\ldots,n_{2x-1},N,n,n_{2x},\ldots} \\
\st{0/1;n_{1},\ldots,n_{2x},\ldots} &
\hf(\nu_{-n_{2x}}-w_{n_{2x}+m}) &
\st{0/1;n_{1},\ldots,n_{2x}+m,\ldots} \\
\st{0/1;n_{1},\ldots,n_{2x-1},\ldots} &
\hf(w_{-n_{2x-1}}-w_{n_{2x-1}+m}) &
\st{0/1;n_{1},\ldots,n_{2x-1}+m,\ldots} \\
\thline
\end{array}
\]
\vspace{-0.25in}
\ca{Breather processes which change the boundary state.}
\label{tab:breproc}
\end{table}
This should be read as implying that any index can have its value
raised, and that a pair of indices can be inserted at any point in the
list, including before the first index and after the last (providing
the resultant state is allowed). Both these tables have been
derived on the basis that, whenever assuming that a pole corresponds to a
bound state leads to a state with the same mass and topological charge as 
one in our minimal spectrum, the assumption is taken to be correct. As
with our
earlier assumption (that, if a pole has another possible explanation, it is
not taken as forming a bound state), this is intuitively reasonable but not
necessarily rigorous. It does, however, lead to consistent results, and
there is no conflict between the two assumptions: we have been unable to
find any alternative explanation for any of the poles listed above. 

It is vital for what follows that, for all the above processes, there
is very little dependence on the existing boundary state. For the
solitons, the topological charge of the state and the value of the
last index in the list are all that matter. Any two states which have
the same topological charge and last index can undergo processes at
the same rapidities to add an index.
Similarly, for the breathers, provided either the relevant two
indices can be added at some point in the list to create an allowed
state, or that the index to be adjusted is present in the list, the
other characteristics of the state are irrelevant.

\goodbreak

\sse{Solitonic pole structure} 
\nobreak
This turns out to be relatively
straightforward. All poles are either of the form \m{\nu_{n}} or
\m{w_{N}}. Looking at a charge 0 state with \m{2k} indices, and
labelling this as 
\m{\underline{x}=(n_{1},n_{2},\ldots,n_{2k})}, we find the results shown
in table \ref{tab:p-0x} for \m{P^{-}_{\st{0;\underline{x}}}(u)}. These poles
come from the \m{a} factors so, for \m{P^{+}}, there is an additional
pole at all \m{\nu}.

\begin{table}[h]
\[
\begin{array}{c|c|c} \thline
\mathbf{Pole} & \mathbf{Order} & \mathbf{Pole} \\
\thline w_{1}\ldots w_{n_{2}-1} &
2k & \nu_{1}\ldots \nu_{n_{1}-1} \\ w_{n_{2}} & 2k-1 & \nu_{n_{1}} \\
w_{n_{2}+1}\ldots w_{n_{4}-1} & 2k-2 & \nu_{n_{1}+1}\ldots \nu_{n_{3}-1} \\
w_{n_{4}} & 2k-3 & \nu_{n_{3}} \\ \vdots & \vdots & \vdots \\
w_{n_{2k-2}+1} \ldots w_{n_{2k}-1} & 2 & \nu_{n_{2k-3}+1} \ldots 
\nu_{n_{2k-1}-1}
\\ w_{n_{2k}} & 1 & \nu_{n_{2k-1}} \\
\thline
\end{array}
\]
\vspace{-0.25in}
\ca{Pole structure for
$\mathbf{P^{-}_{\st{0;\underline{x}}}(u)}$. \mdseries 
An entry of, for example, \m{w_{1}\ldots w_{n_{2}-1}} indicates that
there is a pole of
order \m{2k} at \m{w_{1},w_{2},w_{3},\ldots,w_{n_{2}-1}}.}
\label{tab:p-0x}
\end{table}  

For the charge 1 states, the picture is very similar, and,
considering first the \m{a} factors, we find the pattern given in
table \ref{tab:solch1}  
for a state with \m{2k-1} indices. For \m{P^{-}} there are additional poles at
all \m{w}. (For the charge 0 case, there are poles at \m{w_{x}} for
\m{x \leq 0}, but none of these are in the physical strip.)

\begin{table}[h]
\[
\begin{array}{c|c|c} \thline
\mathbf{Pole} & \mathbf{Order} & \mathbf{Pole} \\
\thline - & 1 & \nu_{-1},\nu_{-2}\ldots \\
- & k & \nu_{0} \\
- & 2k &
\nu_{1}\ldots\nu_{n_{1}-1} \\ - & 2k-1 & \nu_{n_{1}} \\ w_{1} \ldots 
w_{n_{2}-1} &
2k-2 & \nu_{n_{1}+1} \ldots \nu_{n_{3}-1} \\ w_{n_{2}} & 2k-3 & \nu_{n_{3}}
\\ \vdots & \vdots & \vdots \\ w_{n_{2k-4}+1} \ldots w_{n_{2k-2}-1} & 2 &
\nu_{n_{2k-3}+1} \ldots \nu_{n_{2k-1}-1} \\ w_{n_{2k-2}} & 1 &
\nu_{n_{2k-1}} \\
\thline
\end{array}
\]
\vspace{-0.25in}
\ca{Pole structure for
$\mathbf{P^{+}_{\st{1;\underline{x}}}(u)}$.}
\label{tab:solch1}
\end{table}

An important point to note is that, comparing a
general level \m{2k} state \m{\st{0;n_{1},n_{2},\ldots,n_{2k-1},n_{2k}}}
with the level 2 state \m{\st{0;n_{2k-1},n_{2k}}}, we find no additional
poles, though the order of some poles has increased. In the example
above, all level 2 states were explained by diagrams where the
boundary was reduced
either to the vacuum by emission of a breather, or to a first level
excited state by emission of an anti-soliton.
The same processes turn out to be present for any level \m{2k} state
to be reduced to a level \m{2k-1} or \m{2k-2} state. Thus, we might
imagine explaining the poles in the level \m{2k} reflection factor via
similar processes to the ones which explained them in the level 2
factor. At times, however --- as we shall see --- parts of these
processes will need to be replaced by more complex subdiagrams to
allow for the fact that the boundary is
in a higher excited state, explaining the differences in the orders of
the poles. Considering the level 2 processes so far introduced as
``building blocks'', this can be considered as an iterative process: level 4
states can be explained by replacing parts of level 2 processes
with building blocks, while level 6 states can
be explained by similarly replacing parts of level 4 processes with
building blocks, and so on. A generic process of the type we will examine can
therefore be viewed as a cascade of building blocks, each appearing as
a subdiagram of the one before it.
    
A similar argument applies to level \m{2k+1} states and level 3 states,
drawing the same diagrams with all rapidities transformed via \m{\xi
\rightarrow \pi(\la+1)-\xi}. We will concentrate on the charge 0
sector below, and consider a generic level \m{2k} state.

For poles of the form \m{\nu_{n}}, consider \fig{fig:os1}. The
boundary decays to 
the state \m{\st{0;n_{1},n_{2},\ldots,n_{2k-2}}} 
by emission of breather
\m{n_{2k}+n_{2k-1}} at a rapidity of
\m{\hf(\nu_{n_{2k-1}}-w_{n_{2k}})}. This then decays into
breather \m{n_{2k-1}-n} heading towards the boundary at a rapidity of
\m{\hf(w_{-n_{2k-1}}-\nu_{n})} and breather \m{n_{2k}+n} heading
away from the boundary at a rapidity of
\m{\nu_{n}-\left(\hp-\frac{(n_{2k}+n)\pi}{2\la}\right)}. This then
decays to give the outgoing particle and one heading towards the
boundary at a rapidity of \m{w_{n_{2k}}}. For \m{n<n_{2k-1}}, it is
straightforward to check that all these rapidities are within suitable
bounds.

This diagram is na\"{\i}vely third order. However, breather
\m{n_{2k-1}-n}, which is drawn as simply reflecting off the boundary,
in fact has a pole, meaning that the diagram should be treated as
schematic and the appropriate diagram from the next section inserted
instead. In addition, as noted in the discussion of the example, 
the soliton
loop contributes a zero for an incoming anti-soliton through the
Coleman-Thun mechanism. When this is taken into account, we obtain the
correct result.

For \m{\nu_{n_{2k-1}}}, the slightly simpler \fig{fig:fb3} suffices. The
remaining \m{\nu} poles are only present in the soliton reflection
factor, and can be explained by \fig{fig:fb2}, with the boundary
decaying by emitting an anti-soliton at \m{w_{n_{2k}}}, which then
interacts with the incoming soliton to give breather \m{n+n_{2k}},
heading towards the boundary at a rapidity of
\m{\hf(\nu_{n}-w_{n_{2k}})}. Looking ahead again, the interaction
of this breather with the boundary contributes the required zero. For
\m{\nu_{n}<w_{n_{2k}}}, this diagram fails, the breather being created
heading away from the boundary; this is the point when the pole is to
be considered as creating the bound state
\m{\st{1;n_{1},\ldots,n_{2k},n}}.

For the \m{w_{N}} poles, the story is very similar, this time being
based on \fig{fig:fb4} (requiring a 
suitable pole/zero
for \m{B_{N-n_{2k}}} on state
\m{\st{1;n_{1},\ldots,n_{2k-1}}} at
\m{\xl-\hp+\frac{\pi(N+n_{2k}-1)}{2\la}}) for \m{N<n_{2k}} and
\fig{fig:uboun} for \m{n_{2k}}. As noted above, all charge 1 state
poles can be explained by the same mechanisms, with the rapidities
transformed according to \m{\xi \rightarrow \pi(\la+1)-\xi}.

\sse{Breather pole structure}
This is considerably more complicated. However, with a bit of work it
turns out that, for breather \m{n} on the state
\m{\st{0;n_{1},n_{2},\ldots, n_{2k}}}, the pole
structure is as given in table \ref{tab:breps}.
\begin{table}[ht]
\[
\begin{array}{c|c|c} \thline
\mathbf{Pole} & \mathbf{Range} & \mathbf{Pole/zero\ order} \\
\thline \xl-\hp+\frac{\pi(n+2x-1)}{2\la} & n_{2q}<x<n_{2q+2},
n_{2q'}<n+x<n_{2q'+2} & 2(q'-q)+y \\
 & x<0, n_{2q-1}<|x|<n_{2q+1}, n_{2q'} < n- |x| < n_{2q'+2} & 2(q'-q)+y+i
\\
-\xl+\hp+\frac{\pi(n+2x+1)}{2\la} & n_{2q-1}<x<n_{2q+1},
n_{2q'-1}<n+x<n_{2q'+1} & 2(q'-q)+y \\
 & x<0, n_{2q}<|x|<n_{2q+2}, n_{2q'-1}<n-|x|<n_{2q'+1} & 2(q'-q)-i+y \\
 & x<-n, n_{2q} < |x| < n_{2q+2}, n_{2q'} < |x|-n < n_{2q'+2} &
2(q'-q) \\
\xl+\hp+\frac{\pi(n+2x-1)}{2\la} & \mathrm{As\ } \xl -
\hp+\frac{\pi(n+2x+1)}{2\la} \mathrm{\ with\
poles}\leftrightarrow \mathrm{zeroes} & \\
-\xl + \frac{3\pi}{2} + \frac{\pi(n+2x+1)}{2\la} & \mathrm{As\ } -\xl +
\hp + \frac{\pi(n+2x-1)}{2\la} \mathrm{\ with\ poles}\leftrightarrow
\mathrm{zeroes} & \\
\thline
\end{array}
\]
\vspace{-0.25in}
\ca{Breather pole structure for a generic charge 0 state. \mdseries
The variable
$x$ takes integer and half-integer values within the allowed ranges.
An entry in the third column represents a pole of that
order if it is positive, and a zero of appropriate order if it is
negative. (Thus an entry of +1 is a first-order pole, and an entry of
-1 is a first-order zero.) Also, for convenience, \m{y} is 1 if \m{x}
(or \m{|x|}) attains the lower limit, -1 if \m{n+x} (or \m{n-|x|})
attains the lower limit, and zero otherwise, while \m{i} is 1 if \m{x}
is integer, and 0 otherwise.}
\label{tab:breps}
\end{table}

In explaining all this, we can begin with the diagrams found previously.
For the first line, consider an all-breather version of
\fig{fig:brfake}, where the breather decay is chosen to produce breather
\m{n+x-n_{2q'}} on the left, which then binds to the boundary to raise
index \m{n_{2q'}} to \m{n+x}. In some cases, this is not possible, the
appropriate state not being in the spectrum, but, then, we can consider
an all-breather version of \fig{fig:ob1}, where the boundary decays so
as to remove the indices \m{n_{2q'}} and \m{n_{2q'+1}}, with the same
initial breather decay, and the additional breather reflecting from the
boundary contributes a zero. This diagram becomes
possible just as the other fails. In either case, the other breather
from the initial decay (which is drawn as simply reflecting from the
boundary), is breather \m{y=n_{2q'}-x} at rapidity \m{\xl-\hp +
\frac{\pi(y+2x-1)}{2\la}}. This has a pole of order 2 less than the
initial breather. If this order is less than or equal to zero, the
diagram stands as drawn while, otherwise, the simple reflection from the
boundary should be replaced by a repeat of this argument, iterating
until the result is less than or equal to zero. For the next line,
precisely the same argument can be used.

The next three lines can be explained by a similar argument, based on
either increasing the value of index \m{n_{2q'-1}} or removing indices
\m{n_{2q'-1}} and \m{n_{2q'}}.

For \m{\xl+\hp+\frac{\pi(n+2x-1)}{2\la}}, we invoke a similar
process. This time, however, the outer legs have rapidity
\m{\nu_{-(n+x)}} (where \m{-(n+x)} is actually a positive number if
the initial pole is to be in the physical strip), and so we need to
substitute in the explanation of soliton poles of this form from before,
leading, in simple cases, to \fig{fig:ob5}.

Finally, for \m{\frac{3\pi}{2}-\xl+\frac{\pi(n+2x+1)}{2\la}}, we begin with
\fig{fig:ob4}. This time, the reflection factor for the central
soliton always provides a zero, while
the outer soliton has rapidity \m{w_{n+x}}. If \m{n+x=n_{2k}}, the diagram is
as drawn while, otherwise, we need to replace the two outer anti-soliton
legs with the explanation of the appropriate pole in the anti-soliton
factor. The first iteration of this is shown in \fig{fig:ob1}.

\se{Conclusions} 
\label{sec:conclusions}
We now summarise our results. We 
have found that the spectrum of boundary bound states
of the sine-Gordon model with Dirichlet boundary conditions can be
characterised in terms of two ``sectors'', having topological charges
\m{\frac{\beta \varphi_{0}}{2\pi}} and \m{1-\frac{\beta
\varphi_{0}}{2\pi}} (which we labelled
as ``0'' and ``1'' respectively). A boundary state
can be described in an index notation as
\m{\st{c;n_{1},n_{2},\ldots,n_{k}}} for topological charge \m{c}, with
\m{c=0} for
\m{k} even
and \m{c=1} for
\m{k} odd. Such a state can be
created by a succession of alternating solitons and anti-solitons,
beginning with a soliton. The necessary soliton rapidities are of the
form,
\be
\nu_{n}=\xl-\frac{\pi(2n+1)}{2\la}\,,
\ee
while those for the anti-solitons are
\be
w_{m}=\pi-\xl-\frac{\pi(2m-1)}{2\la}\,.
\ee
These are interchanged by the transform \m{\xi
\rightarrow \pi(\la+1) - \xi}. Any such state can be formed, provided
the rapidities involved are monotonically decreasing,
i.e.\ \m{\nu_{n_{1}}>w_{n_{2}}>\nu_{n_{3}}>\ldots}, and its mass is
given by
\ba
m_{n_{1},n_{2},\ldots}&=&
m_s\sin^{2}\left(\frac{\xi-\hp}{2\la}\right)+
\sum_{i \mathrm{\ odd}} m_s\cos(\nu\phup_{n_i})
+ \sum_{j
\mathrm{\ even}} m_s\cos(w\phup_{n_j})\\[4pt]
&=&
m_s\sin^{2}\left(\frac{\xi-\hp}{2\la}\right)+
\sum_{i \mathrm{\ odd}} m_s\cos \left(
\xl - \frac{(2n_{i}+1)\pi}{2\la}\right) - \sum_{j
\mathrm{\ even}} m_s\cos \left( \xl+
\frac{(2n_{j}-1)\pi}{2\la} \right)\,.\nn
\ea
This spectrum is considerably larger than that suggested
in \cite{Skorik}, 
though all the states introduced are required to satisfy our
lemmas. It is worth pointing out that a second part of the analysis
of \cite{Skorik} involved an examination of the (boundary) Bethe
ansatz for a lattice regularisation of the model.
Some of the masses which emerged in the course of that study --
those of the
\m{(n,N)}-strings --- were outwith the spectrum proposed
in \cite{Skorik}, but are now included as the masses of the states
\m{\st{1;0,n,N}}. It remains
to be seen, however,
whether the other masses in our spectrum can be
recovered in the Bethe ansatz approach.

\begin{figure}
\begin{center}
\m{\frac{\xi}{\la+1}}
\parbox{5in}{\includegraphics{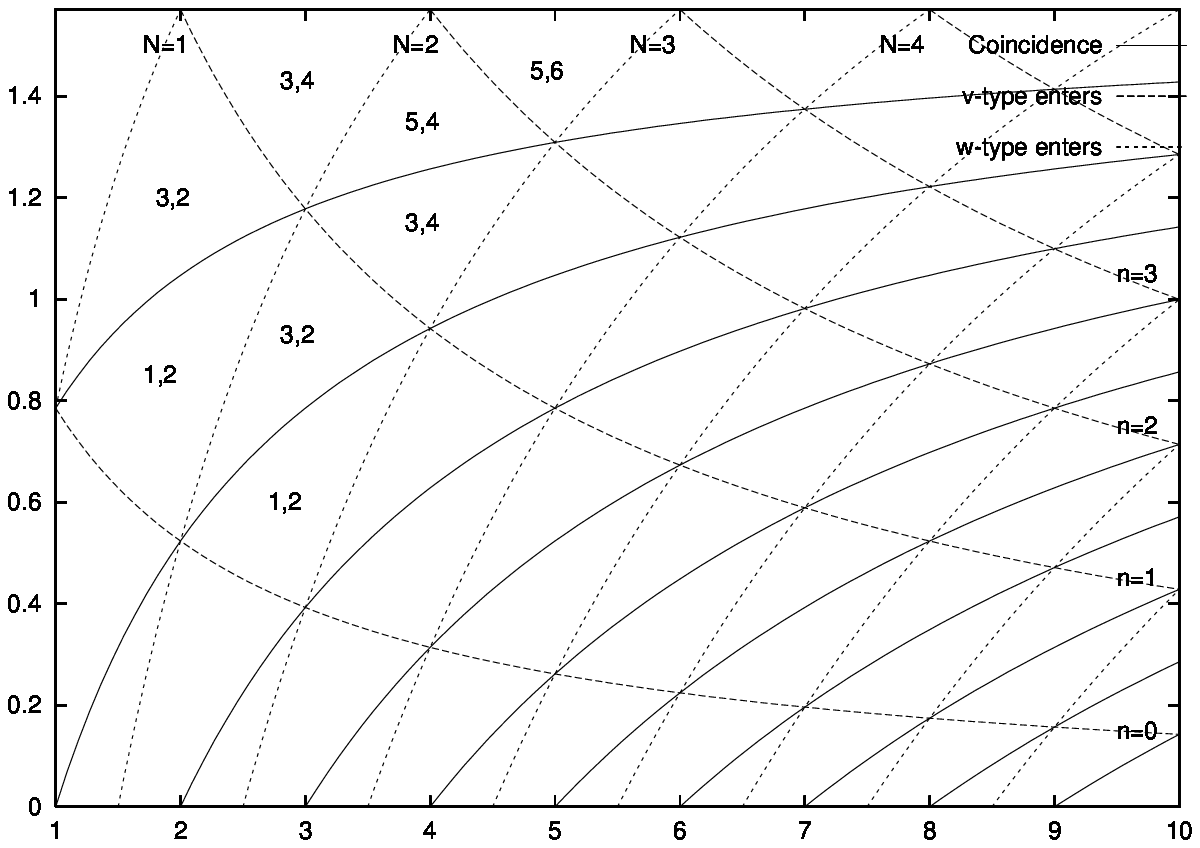}} \\
\m{\la}
\end{center}
\ca{Boundary bound state spectrum size. \mdseries The number of states 
present increases as
$\nu$-type and $w$-type poles enter the physical strip, but changes also
occur as the two sets of poles pass through coincidence: moving in the 
direction of increasing $\la$, the topmost relevant $w$-type pole
passes $\nu_{0}$
and ceases to be relevant, reducing the spectrum. (Notation $x,y$
implies $F(x)$ charge 0 states and $F(y)$ charge 1 states.)}
\label{fig:spectrum}
\end{figure}
The number of states present in the spectrum clearly depends on the
boundary parameters, as illustrated in \fig{fig:spectrum}. It is
convenient to express these in terms of Fibonacci numbers, \m{F(x)},
where \m{F(x)=1,1,2,3,5,\dots} for \m{x=0,1,2,\dots}. If
there are \m{n} \m{\nu}-type poles, and \m{m} relevant\footnote{A
$w$-type pole with a rapidity greater than \m{\nu_{0}} can never be
involved in forming a bound state, and so is not ``relevant'' here.} 
\m{w}-type poles, there are, in general, \m{F(2n)} charge 1 states and
\m{F(2m+1)} charge 0 states. Explicitly, these are given by
\ba
n=\left[\frac{\xi}{\pi}-\hf\right]+1 &\mathrm{and} &
m=\left[\frac{\la}{2}-\frac{\xi}{\pi}+\hf\right] - \left[\la 
- \frac{\xi}{\pi}+\hf\right]\,,
\ea
where the square brackets denote the integer part of the number.
This changes when the two sets of poles
coincide, in which case there are \m{2^{n-1}} states in each sector.

Finally we note that the general method used to
derive the spectrum, via the simple geometrical argument leading to
the two lemmas given in section \ref{sec:colethun}, which can be applied
equally well to any two-dimensional model. Using this to deduce the
existence of as many states as possible led --- in our case --- to the
full spectrum. In other cases, we may not be so fortunate, but, using
it as a starting point, it should make the derivation of the full
spectrum a finite (though possibly lengthy) task.

\smallskip

\noindent
{\bf Acknowledgements --} We are grateful to Aldo Delfino, Subir
Ghoshal, Brett Gibson and Matthias Pillin for help and interesting
discussions. We would especially like to thank Gustav Delius
for his comments on a first draft of this paper.
PM and PED thank the EPSRC for a Research Studentship and 
an Advanced Fellowship respectively, and PED
thanks SPhT Saclay and the Departamento de F\'\i sica
de Part\'\i culas, USC (Santiago de Compostela), for hospitality. The work was
supported in part by a TMR grant of the European Commission, reference
ERBFMRXCT960012.
Finally, we are also grateful to Zoli Bajnok and Gabor Tak\'{a}cs for
pointing out a number of typos in version one of this paper.

\appendix
\setcounter{equation}{0}
\renewcommand{\theequation}{A.\arabic{equation}}
\renewcommand{\thesection}{\textsl{\Alph{section}}}

\se{Infinite products of gamma functions}
\label{app:convergence}
The products which arise in the course of this work are of the form
\be
P(u)=\prod_{l=1}^{\infty}\left[\frac{\g (kl+a-x u)\g(kl+b-xu)}{\g
(kl+c-xu)\g (kl+d-xu)} / (u \rightarrow -u)\right]\,,
\label{eq:app1p}
\ee
Rather than examine this product
directly, we take logs and use the standard formula
\be
\ln \g (z) = z\ln (z) - z - \frac{1}{2}\ln (z) + \ln (\sqrt{2}) +
\frac{1}{12z} + O(z^{-3})
\ee
Assuming that the sum over $l$ and the expansion in $z$ can be exchanged,
potential divergences arise from terms
of the form
\m{\sum_{l=1}^{\infty}\frac{a}{l^{n}}} with \m{a\neq 0} 
and \m{n < 2}. To begin with, we will consider
the terms arising from the block of four terms explicitly shown.

Firstly, there is a contribution of \m{\sum_{l=1}^{\infty}
a+b-c-d} from the \m{z} terms, which
can be set to zero by demanding \m{a+b=c+d}. For the \m{1/12z} terms,
the overall contribution from the
four terms is
\be
\sum_{l=1}^{\infty}
\frac{1}{12}\left(\frac{a-c}{(kl+a-xu)(kl+c-xu)}
+\frac{b-d}{(kl+b-xu)(kl+d-xu)}\right)
\ee
which
can be seen, for \m{a+b=c+d}, to be
of the form \m{1/l^{2}} and hence convergent. 

A similar argument applies to the \m{-\hf \ln (z)} terms, showing they
also provide a convergent contribution. This breaks down when
considering the \m{z\ln(z)} terms, however, and 
their contribution formally reduces to
\be
\sum_{l=1}^{\infty}\Bigl( \frac{cd-ab}{kl} + 
O(l^{-2})\Bigr)~,
\ee
which is divergent unless \m{a=c} or \m{b=c}, both of which are
trivial cases. However, repeating this argument on the other block (with
\m{u \rightarrow -u}) can be seen to yield the same result, allowing
the two divergent terms to cancel, and leaving a product which is
convergent overall.

For comparison with other results, it is useful to write $P(u)$ in other
ways. Firstly, it can be written in terms of Barnes' diperiodic sine
functions using the expansion as given in~\cite{Pillin}:
\ba
S_{2}(x|\w_{1},\w_{2})&=&\exp\left[\frac{(\w_{1}+\w_{2}-2x)\left(\gamma +
\log(2\pi) +
2\log\left(\frac{\w_{1}}{\w_{2}}\right)\right)}{2\w_{1}}\right] \frac{\g
\left( \frac{\w_{1}+\w_{2}-x}{\w_{1}}\right)}{\g
\left(\frac{x}{\w_{1}}\right)} \times \nonumber \\
 & & \prod_{n=1}^{\infty}\left[\frac{\g \left(\frac{\w_{1}+\w_{2}-x +
n\w_{2}} {\w_{1}}\right)}{\g\left( \frac{x+n\w_{2}}{\w_{1}}\right)}
\mathrm{e}^{-\frac{\w_{1}+\w_{2}-2x}{2n\w_{1}}}\left( \frac{n\w_{1}}
{\w_{2}} \right)^{-\frac{\w_{1}+\w_{2}-2x}{\w_{1}}} \right], 
\ea
where \m{\gamma} denotes the Euler constant. For blocks
of the form we are interested in, this simplifies to
\ba
\lefteqn{\frac{S_{2}(x_{1}|\w_{1},\w_{2})S_{2}(x_{2}|\w_{1},\w_{2})}
{S_{2}(x_{3}|\w_{1},\w_{2})S_{2}(x_{4}|\w_{1},\w_{2})}=} \nonumber \\
&& \prod_{n=1}^{\infty} \left[\left\{\frac{\g
\left(\frac{n\w_{2}}{\w_{1}} + \frac{\w_{1}-\w_{2}}{2\w_{1}}-\frac{x'_{1}}{2\w_{1}}\right)\g
\left(\frac{n\w_{2}}{\w_{1}} + \frac{\w_{1}-\w_{2}}{2\w_{1}}-\frac{x'_{2}}{2\w_{1}}\right)}{
\g
\left(\frac{n\w_{2}}{\w_{1}} + \frac{\w_{1}-\w_{2}}{2\w_{1}}-\frac{x'_{3}}{2\w_{1}}\right)\g
\left(\frac{n\w_{2}}{\w_{1}} + \frac{\w_{1}-\w_{2}}{2\w_{1}}-\frac{x'_{4}}{2\w_{1}}\right)}
\right\}/(x'_{m}
\rightarrow -x'_{m})\right]\,,
\ea
(where \m{x'_{m}=x_{m}-\w_{1}-\w_{2}})
provided \m{x_{1}+x_{2}=x_{3}+x_{4}}. Comparing with \re{eq:app1p} we have
\be
P(u)= \frac{S_{2}(\w_{1}(1-a+xu)|\w_{1},\w_{1}k)S_{2}(\w_{1} 
(1-b+xu)|\w_{1},\w_{1}k)}
{S_{2}(\w_{1}(1-c+xu)|\w_{1},\w_{1}k)S_{2}(\w_{1}(1-d+xu)|\w_{1},\w_{1}k)}\,,
\ee
where \m{w_{1}} is arbitrary. In section~\ref{sec:review} we took 
\m{\w_{1}=x^{-1}} for simplicity. The identity
\be
S_{2}(\w_{1}+\w_{2}-x|\w_{1},\w_{2})=\frac{1}{S_{2}(x|\w_{1},\w_{2})}
\ee
was also used. 

These products can also be written in an integral form, through
\be
\log \g (\zeta) = \int_{0}^{\infty} \frac{dx}{x}e^{-x}\left[\zeta-1 +
\frac{e^{-(\zeta-1)x}-1}{1-e^{-x}}\right], \mathrm{Re\ } \zeta>0.
\ee
Since, for the expressions we consider, not all the
\m{\Gamma}-functions have arguments with
positive real part,
it is not possible to give a general formula for \m{P} solely in
these terms. Instead, we give expressions for the reflection
factors. To simplify matters, define
\ba
I^{1}(u)&=&\frac{2\la}{\pi}\int_{-\infty}^{+\infty}dx\cosh \left( \frac{2\la u
x}{\pi}\right)\left[ \frac{\sinh \left( \la -
\frac{2\xi}{\pi}\right)x}{2\sinh x \cosh \la x}\right] \\
I^{2}(u)&=&
\frac{2\la}{\pi}\int_{-\infty}^{+\infty}dx\cosh \left( \frac{2\la u
x}{\pi}\right)\left[\frac{\sinh\left(\frac{2\xi}{\pi}
-2n_{*}-2\right)x}{\sinh x}\right] \\
I^{3}_{n}(u)&=&-\frac{2\la}{\pi}\int_{-\infty}^{+\infty}dx\cosh \left( 
\frac{2\la u x}{\pi}\right)\left[\frac{2\cosh x \sinh \left(\la +1
+2n-\frac{2\xi}{\pi}\right)x}{2\sinh x \cosh \la x}\right] \\
I^{4}_{n}(u)&=&-\frac{2\la}{\pi}\int_{-\infty}^{+\infty}dx\cosh \left( 
\frac{2\la u x}{\pi}\right)\left[\frac{2\cosh x \sinh \left(\frac{2\xi}
{\pi}+2n-\la-1\right)x}{2\sinh x \cosh \la x}\right]
\ea
(where \m{I^{3}_{n}(u)} and \m{I^{4}_{n}(u)} are related to each other 
through \m{\xi \rightarrow \pi(\la+1)-\xi}). The constant \m{n_{*}} is 
the number of \m{\nu}-type poles in the physical strip, which we
recall can be written as
\be
n_{*}=\left[\frac{\xi}{\pi}-\hf\right]\,.
\ee
 
The reflection factors
can then be written as
\ba
-\frac{d}{du}\log \left[\frac{P^{+}_{\st{c;\underline{x}}}(u)}
{R_{0}(u)}\right]&=& 
I^{1}(u)+cI^{2}(u)+\sum_{i \mathrm{\ odd}}I^{3}_{n_{i}}(u) +
\sum_{j \mathrm{\ even}}I^{4}_{n_{j}}(u) \\
-\frac{d}{du}\log \left[\frac{P^{-}_{\st{c;\underline{x}}}(u)}
{R_{0}(u)}\right]&=& 
I^{1}(u)-(1-c)I^{2}(u)+\sum_{i \mathrm{\ odd}}I^{3}_{n_{i}}(u) +
\sum_{j \mathrm{\ even}}I^{4}_{n_{j}}(u)\,,
\ea
for topological charge \m{c} and \m{\underline{x}=(n_{1},n_{2},\ldots, 
n_{2k+c})}. These factors were given in~\cite{Skorik} for the first 
two levels of excited states (the extent of the spectrum they found); 
the above is simply a generalisation of this to the whole spectrum.

\pagebreak

\se{On-shell diagrams}
In this appendix we collect together some of the on-shell diagrams
used in the main body of the paper.
All boundaries are initially in the state
\m{\st{n_{1},n_{2},\ldots,n_{2k}}}, where \m{k} can be any integer, and
we have suppressed the topological charge index (which is
zero). Analogous processes for charge 1 states can be found by
applying the transformation \m{\xi \rightarrow \pi(\la+1)-\xi} to all
rapidities shown. 

In addition, where the boundary is shown decaying
through emission of a breather, only the process where this removes
the last two indices is given. Similar processes always exist to
remove any other adjacent pair of indices, or to simply modify an
index; see section \ref{sec:refmin} for the appropriate breather
boundary vertices.

\vskip 2cm

\begin{figure}[!ht]
\begin{center}
\parbox{3.4in}{
\unitlength 1.00mm
\linethickness{0.4pt}
\begin{picture}(48.00,55.00)(0,2)
\put(46.00,3.00){\rule{2.00\unitlength}{52.00\unitlength}}
\color{red}
\multiput(45.67,53.11)(0,1){2}{\circle*{0.25}}
\multiput(45.78,4.00)(0,1){2}{\circle*{0.25}}
\multiput(45.67,6.00)(0,2){24}{\circle*{0.25}}
\color{blue}
\multiput(37.33,13.44)(1,-1){9}{\circle*{0.25}}
\multiput(37.22,44.67)(1,1){9}{\circle*{0.25}}
\color{green}
\put(36.78,44.22){\line(3,-5){9.07}}
\put(36.78,14.00){\line(3,5){9.07}}
\put(36.78,14.00){\line(-2,-1){24.44}}
\put(36.78,44.22){\line(-2,1){24.44}}
\color{deepred}
\multiput(36.78,11.00)(0,1){7}{\circle*{0.25}}
\qbezier(36.78,11.00)(34.00,10.00)(32.78,12.00)
\put(32.78,14){\makebox(0,0)[rc]{$w_{n_{2k}-n}$}}
\qbezier(42.22,49.67)(43.44,47.67)(45.56,48.33)
\qbezier(44.11,32.11)(44.56,33.44)(45.89,32.89)
\put(35.00,32.18){\makebox(0,0)[lc]{$w_{n_{2k}}$}}
\put(50.00,48.67){\makebox(0,0)[lc]{$\frac{\nu_{n_{2k-1}}-w_{n_{2k}}}{2}$}}
\put(50.00,30.00){\makebox(0,0)[lc]{$\st{n_{1},\ldots,n_{2k-2}}$}}
\put(40.00,8.00){\makebox(0,0)[cc]{$n$}}
\put(40.00,50.00){\makebox(0,0)[cc]{$n$}}
\color{black}
\end{picture}
\ca{Soliton, breather boundary decay}
\label{fig:fb1} } \
\parbox{3.4in}{
\unitlength 1.00mm
\linethickness{0.4pt}
\begin{picture}(48.00,55.00)(0,2)
\color{blue}
\multiput(46.18,9.44)(-1,2){6}{\circle*{0.25}}
\multiput(46.18,49.67)(-1,-2){6}{\circle*{0.25}}
\color{red}
\multiput(45.67,49.11)(0,1){6}{\circle*{0.25}}
\multiput(45.78,4.00)(0,1){6}{\circle*{0.25}}
\multiput(45.67,10.00)(0,2){10}{\circle*{0.25}}
\multiput(45.67,32.00)(0,2){7}{\circle*{0.25}}
\color{green}
\put(40.96,20.00){\line(1,2){5.03}}
\put(40.96,19.89){\line(-5,3){28.11}}
\put(40.96,40.11){\line(1,-2){5.03}}
\put(40.96,40.22){\line(-5,-3){28.11}}
\color{deepred}
\multiput(23.96,24.99)(0,1){11}{\circle*{0.25}}
\qbezier(20.96,28.22)(21.74,26.56)(23.96,27.00)
\qbezier(44.05,45.78)(44.41,44.11)(46.52,44.67)
\put(20.96,29.89){\makebox(0,0)[rc]{$\nu_{n-n_{2k}}$}}
\put(50.00,23.18){\makebox(0,0)[lc]{$w_{n_{2k}}$}}
\put(50.00,45.67){\makebox(0,0)[lc]{$\frac{\nu_{n_{2k-1}}-w_{n_{2k}}}{2}$}}
\qbezier(43.33,24.56)(44.00,23.11)(46.00,23.78)
\qbezier(36.89,22.44)(39.11,24.67)(42.78,23.67)
\put(36.67,19.89){\vector(3,2){3.67}}
\put(30.89,18.78){\makebox(0,0)[cc]{$\pi-\frac{n\pi}{\la}$}}
\put(40.67,15.11){\makebox(0,0)[cc]{$n$}}
\put(41.11,45.22){\makebox(0,0)[cc]{$n$}}
\put(50.00,30.00){\makebox(0,0)[lc]{$\st{n_{1},\ldots,n_{2k-2}}$}}
\color{black}
\put(46.00,3.00){\rule{2.00\unitlength}{52.00\unitlength}}
\end{picture}
\ca{Incoming soliton crossed}
\label{fig:fb3} }
\end{center}
\vskip 1.5cm
\end{figure}

\begin{figure}[!ht]
\begin{center}
\parbox{3.4in}{
\unitlength 1.00mm
\linethickness{0.4pt}
\begin{picture}(48.00,55.24)(0,2)
\put(46.00,3.00){\rule{2.00\unitlength}{52.00\unitlength}}
\color{red}
\multiput(45.67,52.11)(0,1){3}{\circle*{0.25}}
\multiput(45.78,4.00)(0,1){3}{\circle*{0.25}}
\multiput(45.78,8.00)(0,2){22}{\circle*{0.25}}
\color{green}
\put(46.00,5.67){\line(-1,1){7.78}}
\put(46.00,51.89){\line(-1,-1){7.78}}
\put(38.22,13.44){\line(-6,-1){24.78}}
\put(38.22,44.11){\line(-6,1){24.78}}
\color{blue}
\multiput(38.22,13.44)(0.5,1){16}{\circle*{0.25}}
\multiput(38.22,44.11)(0.5,-1){16}{\circle*{0.25}}
\color{deepred}
\multiput(38.22,48.67)(0,-1){10}{\circle*{0.25}}
\qbezier(34.89,44.67)(35.00,47.56)(38.22,47.78)
\put(34.00,47.78){\makebox(0,0)[rc]{$\nu_{n-n_{2k}}$}}
\qbezier(42.44,9.22)(43.00,11.89)(45.89,11.78)
\qbezier(34.00,12.78)(36.33,9.78)(40.44,11.22)
\put(35.44,15.56){\vector(1,-4){1}}
\put(31.89,18.11){\makebox(0,0)[cc]{$\pi-\frac{n\pi}{\la}$}}
\put(40.00,22.78){\makebox(0,0)[cc]{$n$}}
\put(40.00,33.78){\makebox(0,0)[cc]{$n$}}
\qbezier(46.00,24.44)(45.00,21.44)(44.00,24.44)
\put(50.00,30.00){\makebox(0,0)[lc]{$\st{n_{1},\ldots,n_{2k-1}}$}}
\put(50.00,9.56){\makebox(0,0)[lc]{$w_{n_{2k}}$}}
\put(50.00,23.00){\makebox(0,0)[lc]{$\frac{\nu_{n}-w_{2n_{2k}}}{2}$}}
\color{black}
\end{picture}
\ca{Soliton, soliton boundary decay}
\label{fig:fb2} } \
\parbox{3.4in}{
\unitlength 1.00mm
\linethickness{0.4pt}
\begin{picture}(48.00,55.00)(0,2)
\put(46.00,3.00){\rule{2.00\unitlength}{52.00\unitlength}}
\color{blue}
\multiput(39.22,22.33)(1,1){7}{\circle*{0.25}}
\multiput(39.22,35.44)(1,-1){7}{\circle*{0.25}}
\color{green}
\put(4.33,18.00){\line(2,1){34.78}}
\put(39.22,22.44){\line(-2,1){34.78}}
\put(39.11,35.44){\line(1,2){6.61}}
\put(45.89,9.22){\line(-1,2){6.61}}
\color{deepred}
\qbezier(43.67,13.89)(44.11,15.78)(45.89,14.56)
\qbezier(20.89,26.22)(21.89,23.67)(26.11,24.67)
\qbezier(34.56,33.22)(35.67,38.56)(41.33,39.89)
\put(30.67,40.00){\makebox(0,0)[cc]{$\pi - \frac{n\pi}{\lambda}$}}
\put(20.89,27.22){\makebox(0,0)[rc]{$w_{n_{2k}-n}$}}
\multiput(26.11,22.33)(0,1){14}{\circle*{0.25}}
\qbezier(41.89,24.78)(43.00,22.22)(46.00,22.78)
\put(50.00,30.00){\makebox(0,0)[lc]{$\st{n_{1},\ldots,n_{2k-1}}$}}
\put(40.00,25.78){\makebox(0,0)[cc]{$n$}}
\put(40.00,31.78){\makebox(0,0)[cc]{$n$}}
\put(50.00,12.56){\makebox(0,0)[lc]{$w_{n_{2k}}$}}
\put(50.00,23.00){\makebox(0,0)[lc]{$\frac{\nu_{n}-w_{2n_{2k}}}{2}$}}
\color{red}
\multiput(45.67,49.00)(0,1){6}{\circle*{0.25}}
\multiput(45.67,4.00)(0,1){6}{\circle*{0.25}}
\multiput(45.67,11.00)(0,2){19}{\circle*{0.25}}
\color{black}
\end{picture}
\ca{Incoming soliton crossed}
\label{fig:fb4} }
\end{center}
\vskip 1.5cm
\end{figure}

\begin{figure}[!ht]
\begin{center}
\parbox{3.4in}{
\unitlength 1.00mm
\linethickness{0.4pt}
\begin{picture}(48.00,55.00)(0,2)
\put(46.00,3.00){\rule{2.00\unitlength}{52.00\unitlength}}
\color{blue}
\multiput(24.89,4.45)(1,1){11}{\circle*{0.25}}
\multiput(24.89,53.77)(1,-1){11}{\circle*{0.25}}
\color{green}
\put(34.78,43.78){\line(3,-1){11}}
\put(34.78,43.78){\line(3,-4){11}}
\put(34.78,14.44){\line(3,1){11}}
\put(34.78,14.44){\line(3,4){11}}
\color{red}
\multiput(45.78,18.22)(0,1){22}{\circle*{0.25}}
\color{deepred}
\qbezier(39.78,42.00)(39.67,40.22)(37.89,39.78)
\qbezier(42.44,16.89)(43.67,15.00)(45.89,15.22)
\put(28.88,39.78){\makebox(0,0)[cc]{$\pi-\frac{n\pi}{2\lambda}$}}
\put(50.00,15.78){\makebox(0,0)[lc]{$\nu_{m}$}}
\qbezier(42.33,24.44)(43.22,23.11)(46.00,23.44)
\put(50.00,23.44){\makebox(0,0)[lc]{$\nu_{m-n}$}}
\put(50.00,30.00){\makebox(0,0)[lc]{$\st{n_{1},\ldots,n_{2k},m}$}}
\put(28.78,10.89){\makebox(0,0)[cc]{$n$}}
\put(28.78,47.89){\makebox(0,0)[cc]{$n$}}
\multiput(34.78,11.44)(0,1){7}{\circle*{0.25}}
\qbezier(32.78,12.44)(33.00,10.00)(34.78,11.44)
\put(32.78,14.44){\makebox(0,0)[rc]{$\frac{\nu_{2m}-w_{n}}{2}$}}
\color{black}
\end{picture}
\ca{Breather, soliton bound state}
\label{fig:brfake} } \
\parbox{3.4in}{
\unitlength 1.00mm
\linethickness{0.4pt}
\begin{picture}(48.00,55.00)(0,2)
\color{deepred}
\multiput(45.67,3.44)(0,1){52}{\circle*{0.25}}
\qbezier(41.44,9.22)(43.00,11.22)(45.89,10.44)
\qbezier(40.44,30.78)(42.67,33.00)(45.89,32.44)
\put(50.00,8.44){\makebox(0,0)[lc]{$w_{n_{2k}}$}}
\put(50.00,31.00){\makebox(0,0)[lc]{$w_{n_{2k}+n}+\pi$}}
\qbezier(21.89,24.67)(25.11,24.56)(24.44,21.78)
\put(23.89,18.44){\vector(-1,2){1.83}}
\put(23.89,17.11){\makebox(0,0)[cc]{$\pi-\frac{n\pi}{2\la}$}}
\put(9.78,15.89){\makebox(0,0)[cc]{$n$}}
\put(9.78,41.89){\makebox(0,0)[cc]{$n$}}
\put(50.00,25.00){\makebox(0,0)[lc]{$\st{n_{1},\ldots,n_{2k-1}}$}}
\multiput(16.78,16.22)(0,1){7}{\circle*{0.25}}
\qbezier(14.12,17.22)(15.00,16.22)(16.78,17.22)
\put(14.00,19.5){\makebox(0,0)[rc]{$\pi+\frac{w_{2n_{2k}}-\nu_{n}}{2}$}}
\color{green}
\put(45.89,3.67){\line(-5,6){29.07}}
\put(45.89,54.11){\line(-5,-6){29.07}}
\put(16.78,38.56){\line(3,-1){29.11}}
\put(16.78,19.22){\line(3,1){29.11}}
\color{blue}
\multiput(16.78,38.56)(-1.33,1){13}{\circle*{0.25}}
\multiput(16.78,19.22)(-1.33,-1){13}{\circle*{0.25}}
\color{black}
\put(46.00,3.00){\rule{2.00\unitlength}{52.00\unitlength}}
\end{picture}
\ca{Outgoing soliton crossed}
\label{fig:ob3} }
\end{center}
\vskip 1cm
\end{figure}
 
\begin{figure}[!ht]
\begin{center}
\parbox{3.4in}{
\unitlength 1.00mm
\linethickness{0.4pt}
\begin{picture}(48.00,55.24)
\put(46.00,3.00){\rule{2.00\unitlength}{52.00\unitlength}}
\color{red}
\multiput(45.67,52.11)(0,1){3}{\circle*{0.25}}
\multiput(45.78,4.00)(0,1){3}{\circle*{0.25}}
\multiput(45.78,8.00)(0,2){22}{\circle*{0.25}}
\color{green}
\put(46.00,5.67){\line(-1,1){7.78}}
\put(38.22,13.44){\line(1,2){7.67}}
\put(46.00,51.89){\line(-1,-1){7.78}}
\put(38.22,44.11){\line(1,-2){7.67}}
\color{blue}
\multiput(38.22,13.56)(-1,-0.33){24}{\circle*{0.25}}
\multiput(38.22,44.11)(-1,0.33){24}{\circle*{0.25}}
\color{deepred}
\qbezier(42.44,9.22)(43.00,11.89)(45.89,11.78)
\put(50.00,9.56){\makebox(0,0)[lc]{$w_{n_{2k}}$}}
\qbezier(42.22,36.22)(43.33,37.78)(45.89,37.44)
\put(50.00,35.67){\makebox(0,0)[lc]{$-w_{n_{2k}+n}$}}
\qbezier(41.44,47.22)(43.33,43.89)(40.89,39.11)
\put(29.44,40.67){\makebox(0,0)[cc]{$\pi - \frac{n\pi}{\la}$}}
\put(36.44,41.67){\vector(4,1){4}}
\put(24.67,47.00){\makebox(0,0)[cc]{$n$}}
\put(24.56,10.11){\makebox(0,0)[cc]{$n$}}
\put(50.00,30.00){\makebox(0,0)[lc]{$\st{n_{1},\ldots,n_{2k-1}}$}}
\multiput(38.22,17.06)(0,-1){8}{\circle*{0.25}}
\qbezier(34.56,12.39)(35.45,10.28)(38.22,10.72)
\put(34.00,14.44){\makebox(0,0)[rc]{$\frac{\nu_{n}-w_{2n_{2k}}}{2}$}}
\color{black}
\end{picture}
\ca{Breather, soliton boundary decay}
\label{fig:ob6} } \
\parbox{3.4in}{
\unitlength 1.00mm
\linethickness{0.4pt}
\begin{picture}(48.00,55.00)
\put(46.00,3.00){\rule{2.00\unitlength}{52.00\unitlength}}
\color{green}
\put(33.22,35.89){\line(5,-3){12.78}}
\put(33.33,35.78){\line(3,4){12.67}}
\put(33.22,20.56){\line(5,3){12.78}}
\put(33.33,20.67){\line(3,-4){12.67}}
\color{blue}
\multiput(33.44,20.56)(-1,0.33){30}{\circle*{0.25}}
\multiput(33.44,35.89)(-1,-0.33){30}{\circle*{0.25}}
\color{deepred}
\multiput(10.44,31.78)(0,-1){8}{\circle*{0.25}}
\qbezier(6.78,26.89)(7.67,24.78)(10.44,25.22)
\put(9.11,21.44){\makebox(0,0)[cc]{$\pi+\frac{w_{2n_{2k}}-\nu_{n}}{2}$}}
\qbezier(41.00,46.11)(43.22,44.11)(45.89,44.22)
\put(50.00,47.00){\makebox(0,0)[lc]{$w_{n_{2k}}$}}
\qbezier(40.11,24.78)(41.67,22.22)(45.89,22.67)
\put(50.00,24.67){\makebox(0,0)[lc]{$-w_{n_{2k}+n}$}}
\qbezier(37.22,23.00)(39.33,18.89)(36.56,16.44)
\put(30.78,17.11){\vector(2,1){5.33}}
\put(24.78,16.11){\makebox(0,0)[cc]{$\pi - \frac{n\pi}{\la}$}}
\put(21.22,33.44){\makebox(0,0)[cc]{$n$}}
\put(21.11,23.00){\makebox(0,0)[cc]{$n$}}
\put(50.00,30.00){\makebox(0,0)[lc]{$\st{n_{1},\ldots,n_{2k-1}}$}}
\color{black}
\end{picture}
\ca{Incoming breather crossed}
\label{fig:ob4} }
\vskip 1cm
\end{center}
\end{figure}

\begin{figure}[!ht]
\begin{center}
\parbox{3.4in}{
\unitlength 0.80mm
\linethickness{0.4pt}
\begin{picture}(48.00,69.25)(-12,-7.5)
\put(46.00,-3.00){\rule{2.50\unitlength}{65.00\unitlength}}
\color{red}
\multiput(45.67,-2.00)(0,2){32}{\circle*{0.25}}
\color{blue}
\multiput(18.44,58.94)(-2,1){7}{\circle*{0,25}}
\multiput(18.44,4.72)(-2,-1){7}{\circle*{0.25}}
\multiput(35.78,8.06)(1,-1){11}{\circle*{0.25}}
\multiput(40.44,54.94)(1,1){6}{\circle*{0.25}}
\color{green}
\put(45.89,31.84){\line(-1,1){27.34}}
\put(45.89,31.84){\line(-1,-1){27.34}}
\put(18.66,58.94){\line(5,-1){21.56}}
\put(40.22,54.86){\line(1,-3){5.60}}
\put(45.89,38.50){\line(-1,-3){10.14}}
\put(18.88,4.94){\line(5,1){16.66}}
\color{deepred}
\multiput(18.44,55.94)(0,1){7}{\circle*{0.25}}
\qbezier(18.44,61.94)(16.00,63.00)(14.44,60.94)
\put(14.44,58.94){\makebox(0,0)[cc]{$u$}}
\put(50.00,58.94){\makebox(0,0)[lc]{$u=\frac{w_{0}-\nu_{2n_{2k-1}+n}}{2}$}}
\qbezier(41.00,27.00)(43.00,24.33)(46.00,26.00)
\put(50.00,30.00){\makebox(0,0)[lc]{$-\nu_{n_{2k-1}+n}$}}
\qbezier(43.22,46.00)(43.89,47.78)(45.89,46.78)
\put(50.00,43.44){\makebox(0,0)[lc]{$w_{n_{2k}}$}}
\qbezier(23.89,9.78)(27.67,9.44)(26.22,6.33)
\put(19.89,10.44){\vector(4,-3){4.44}}
\put(13.00,12.11){\makebox(0,0)[cc]{$\pi-\frac{n\pi}{\la}$}}
\qbezier(30.56,7.22)(31.56,11.78)(37.56,13.67)
\put(31.56,4.56){\vector(1,2){3.17}}
\put(26.56,1.67){\makebox(0,0)[cc]{$\pi-\frac{l\pi}{\la}$}}
\qbezier(40.00,4.11)(42.22,7.89)(45.89,6.78)
\put(50.00,9.78){\makebox(0,0)[lc]{$\frac{\nu_{n_{2k-1}}-w_{n_{2k}}}{2}$}}
\put(37.56,2.22){\makebox(0,0)[cc]{$l$}}
\put(50.00,52.22){\makebox(0,0)[lc]{$l=n_{2k}+n_{2k-1}$}}
\put(11.67,3.00){\makebox(0,0)[cc]{$n$}}
\put(50.00,23.56){\makebox(0,0)[lc]{$\st{n_{1},\ldots,n_{2k-2}}$}}
\color{black}
\end{picture}
\ca{As \ref{fig:brfake}, outer legs replaced by \ref{fig:fb1}}
\label{fig:ob5} } \
\parbox{3.4in}{
\unitlength 1.00mm
\linethickness{0.4pt}
\begin{picture}(48.00,55.00)
\put(46.00,3.00){\rule{2.00\unitlength}{52.00\unitlength}}
\color{red}
\multiput(45.67,3.44)(0,1){52}{\circle*{0.25}}
\color{green}
\put(31.22,41.77){\line(5,6){6.94}}
\put(45.89,17.77){\line(-1,-4){1.67}}
\put(38.00,49.78){\line(1,-4){8.00}}
\put(31.22,41.78){\line(4,-1){14.67}}
\put(25.56,32.78){\line(5,-6){18.52}}
\put(45.89,17.89){\line(-1,-4){1.81}}
\put(25.56,32.78){\line(4,1){20.44}}
\color{blue}
\multiput(2.11,42.23)(1.25,-0.5){19}{\circle*{0.25}}
\multiput(31.11,41.67)(-1.25,-0.5){24}{\circle*{0.25}}
\multiput(44.00,10.56)(0.25,-1){7}{\circle*{0.25}}
\multiput(38.00,49.78)(0.25,1){5}{\circle*{0.25}}
\color{deepred}
\multiput(17.30,39.00)(0,-1){7}{\circle*{0.25}}
\qbezier(17.3,34.00)(16.00,33.00)(14.8,35.00)
\put(13.00,36.00){\makebox(0,0)[cc]{$u$}}
\put(8.00,42.00){\makebox(0,0)[cc]{$n$}}
\put(8.00,30.00){\makebox(0,0)[cc]{$n$}}
\put(50.00,51.94){\makebox(0,0)[lc]{$u=\pi+\frac{\nu_{2n_{2k-1}+n}
-w_{0}}{2}$}}
\put(50.00,45.94){\makebox(0,0)[lc]{$l=n_{2k}+n_{2k-1}$}}
\qbezier(44.33,9.22)(44.67,11.00)(46.00,10.00)
\put(50.00,8.00){\makebox(0,0)[lc]{$\frac{\nu_{n_{2k-1}}-w_{n_{2k}}}{2}$}}
\qbezier(44.00,26.33)(44.44,28.00)(45.89,27.00)
\put(50.00,23.78){\makebox(0,0)[lc]{$w_{n_{2k}}$}}
\qbezier(42.00,39.11)(42.78,41.89)(45.89,42.00)
\put(50.00,40.11){\makebox(0,0)[lc]{$-\nu_{n_{2k-1}+n}$}}
\qbezier(31.56,34.33)(32.33,30.44)(30.44,27.00)
\put(16.44,28.44){\makebox(0,0)[cc]{$\pi - \frac{n\pi}{2\la}$}}
\put(22.44,30.44){\vector(4,1){6}}
\qbezier(33.89,45.22)(36.56,43.11)(39.22,44.89)
\put(29.44,48.33){\vector(4,-1){7.22}}
\put(23.33,47.56){\makebox(0,0)[cc]{$\pi - \frac{n\pi}{2\la}$}}
\put(37.33,53.00){\makebox(0,0)[cc]{$l$}}
\put(43.44,6.22){\makebox(0,0)[cc]{$l$}}
\put(50.00,30.00){\makebox(0,0)[lc]{$\st{n_{1},\ldots,n_{2k-2}}$}}
\color{black}
\end{picture}
\ca{As \ref{fig:ob4}, outer legs replaced by \ref{fig:fb3}}
\label{fig:ob1} }
\vskip 1cm
\end{center}
\end{figure}

\begin{figure}[!ht]
\begin{center}
\parbox{3.4in}{
\unitlength 1.00mm
\linethickness{0.4pt}
\begin{picture}(48.00,55.00)
\color{red}
\multiput(45.67,3.44)(0,1){52}{\circle*{0.25}}
\color{green}
\put(34.22,30.11){\line(5,-6){4.44}}
\put(45.89,39.45){\line(-1,2){2.83}}
\put(38.78,24.89){\line(1,2){7.67}}
\put(34.22,30.11){\line(5,-2){11.67}}
\put(45.89,25.33){\line(-5,-2){27.78}}
\put(43.00,45.11){\line(-4,-5){24.71}}
\color{blue}
\multiput(18.22,14.22)(-1.5,-1){8}{\circle*{0.25}}
\multiput(34.22,30.11)(-1.5,1){16}{\circle*{0.25}}
\multiput(43.00,45.23)(0.5,1.5){6}{\circle*{0.25}}
\multiput(38.78,24.67)(0.5,-1.5){14}{\circle*{0.25}}
\color{deepred}
\multiput(18.22,11.22)(0,1){7}{\circle*{0.25}}
\qbezier(18.22,12.22)(16.50,11.00)(15.22,12.22)
\put(16.5,10.00){\makebox(0,0)[cc]{$u$}}
\put(50.00,51.94){\makebox(0,0)[lc]{$u=\pi+\frac{\nu_{2n_{2k-1}}-w_{0}}{2}$}}
\put(50.00,45.94){\makebox(0,0)[lc]{$a=2\pi-\frac{\pi(n_{2k-1}
+n_{2k}+n)}{\la}$}}
\put(50.00,39.94){\makebox(0,0)[lc]{$l=n_{2k}+n_{2k-1}$}}
\qbezier(43.78,10.12)(45.00,11.67)(45.89,11.00)
\qbezier(43.56,24.44)(44.00,23.11)(45.89,23.00)
\qbezier(43.78,34.45)(44.56,33.23)(45.89,33.67)
\qbezier(39.44,26.44)(41.00,25.67)(41.66,27.11)
\put(50.00,9.67){\makebox(0,0)[lc]{$\frac{\nu_{n_{2k-1}}-w_{n_{2k}}}{2}$}}
\put(50.00,22.22){\makebox(0,0)[lc]{$\pi+\nu_{n_{2k-1}+n}$}}
\put(50.00,34.45){\makebox(0,0)[lc]{$w_{n_{2k}}$}}
\put(39.00,31.00){\makebox(0,0)[cc]{$a$}}
\put(39.33,30.11){\vector(1,-3){1.2}}
\qbezier(24.00,21.22)(27.56,20.56)(26.78,17.67)
\put(27.11,13.78){\vector(-2,3){3}}
\put(27.00,11.67){\makebox(0,0)[cc]{$\pi-\frac{n\pi}{\la}$}}
\put(12.33,12.56){\makebox(0,0)[cc]{$n$}}
\put(22.22,36.00){\makebox(0,0)[cc]{$n$}}
\put(43.00,8.00){\makebox(0,0)[cc]{$l$}}
\put(43.00,48.00){\makebox(0,0)[cc]{$l$}}
\put(50.00,29.00){\makebox(0,0)[lc]{$\st{n_{1},\ldots,n_{2k-2}}$}}
\color{black}
\put(46.00,3.00){\rule{2.00\unitlength}{52.00\unitlength}}
\end{picture}
\ca{As \ref{fig:ob3}, outer legs replaced by \ref{fig:fb3}}
\label{fig:ob2} } \
\parbox{3.4in}{
\unitlength 1.00mm
\linethickness{0.4pt}
\begin{picture}(48.00,55.00)
\color{red}
\multiput(45.67,4.00)(0,1){10}{\circle*{0.25}}
\multiput(45.67,15.00)(0,2){16}{\circle*{0.25}}
\multiput(45.67,46.00)(0,1){9}{\circle*{0.25}}
\color{blue}
\multiput(32.22,43.22)(-2,1){8}{\circle*{0,25}}
\multiput(32.22,16.11)(-2,-1){8}{\circle*{0.25}}
\multiput(43.11,41.18)(0.5,-1.5){6}{\circle*{0.25}}
\multiput(45.89,33.00)(-0.5,-1.5){11}{\circle*{0.25}}
\color{green}
\put(45.89,29.67){\line(-1,1){13.67}}
\put(45.89,29.67){\line(-1,-1){13.67}}
\put(40.89,17.78){\line(1,-1){5}}
\put(43.22,41.22){\line(1,1){4}}
\put(32.33,43.33){\line(5,-1){10.78}}
\put(32.44,16.22){\line(5,1){8.33}}
\color{deepred}
\multiput(32.22,40.22)(0,1){7}{\circle*{0.25}}
\qbezier(32.22,45.22)(30.5,46.50)(29.22,44.72)
\put(31.00,47.00){\makebox(0,0)[cc]{$u$}}
\qbezier(41.44,17.22)(43.00,20.44)(46.00,20.44)
\put(50.00,17.67){\makebox(0,0)[lc]{$w_{n_{2k}}$}}
\qbezier(44.00,40.22)(44.89,41.11)(45.89,41.11)
\put(49.33,39.22){\vector(-1,0){5.11}}
\put(50.78,39.22){\makebox(0,0)[lc]{$\frac{\nu_{l}-w_{2n_{2k}}}{2}$}}
\qbezier(41.56,25.33)(42.44,23.78)(46.00,23.89)
\put(49.22,26.33){\vector(-1,0){5.44}}
\put(51.22,26.33){\makebox(0,0)[lc]{$-w_{n_{2k}+n-l}$}}
\qbezier(36.67,17.11)(39.22,15.44)(42.78,15.89)
\put(38.22,12.56){\vector(1,2){2.06}}
\put(34.33,9.11){\makebox(0,0)[cc]{$\pi-\frac{l\pi}{\la}$}}
\put(35.67,24.11){\vector(2,-1){4.44}}
\put(33.22,24.78){\makebox(0,0)[cc]{$l$}}
\qbezier(36.56,20.44)(39.22,20.11)(38.44,17.56)
\put(32.11,19.22){\vector(1,0){5.44}}
\put(25.89,20.00){\makebox(0,0)[cc]{$\pi-\frac{n\pi}{\la}$}}
\put(23.56,13.89){\makebox(0,0)[cc]{$n$}}
\put(23.56,44.89){\makebox(0,0)[cc]{$n$}}
\put(50.00,32.00){\makebox(0,0)[lc]{$\st{n_{1},\ldots,n_{2k-1}}$}}
\put(50.00,51.94){\makebox(0,0)[lc]{$u=\frac{\nu_{2l}-w_{2n_{2k}+n}}{2}$}}
\color{black}
\put(46.00,3.00){\rule{2.00\unitlength}{52.00\unitlength}}
\end{picture}
\ca{As \ref{fig:brfake}, outer legs replaced by \ref{fig:fb2}}
\label{fig:ob7} }
\vskip 1cm
\end{center}
\end{figure}

\begin{figure}[!ht]
\begin{center}
\parbox{5in}{
\centering
\unitlength 1.00mm
\linethickness{0.4pt}
\begin{picture}(65.00,55.57)(7,0)
\put(46.00,3.00){\rule{2.00\unitlength}{52.00\unitlength}}
\color{red}
\multiput(45.67,4.00)(0,2){26}{\circle*{0.25}}
\color{green}
\put(32.92,10.23){\line(-5,1){25.26}}
\put(25.44,38.67){\line(-5,-1){15.00}}
\put(38.37,41.33){\line(-5,-1){25.26}}
\put(45.89,29.89){\line(-2,-3){13.04}}
\put(45.89,29.89){\line(-2,-3){12.07}}
\put(45.89,29.78){\line(-2,-3){10.15}}
\put(46.00,29.89){\line(-2,3){7.63}}
\color{blue}
\multiput(34.67,6.56)(-0.5,1){5}{\circle*{0.25}}
\multiput(38.33,41.33)(0.5,0.83){7}{\circle*{0.25}}
\multiput(41.33,46.33)(0.25,1.5){8}{\circle*{0.25}}
\multiput(34.72,6.66)(0.25,-1.25){6}{\circle*{0.25}}
\multiput(45.89,34.78)(-0.5,-1.25){21}{\circle*{0.25}}
\multiput(45.89,34.89)(-0.5,-1.25){23}{\circle*{0.25}}
\multiput(41.33,46.33)(0.5,-1.25){10}{\circle*{0.25}}
\multiput(45.89,34.78)(-0.5,-1.25){18}{\circle*{0.25}}
\color{deepred}
\qbezier(32.89,40.22)(35.33,36.00)(40.56,38.00)
\put(29.11,35.44){\makebox(0,0)[cc]{$\pi-\frac{n\pi}{2\la}$}}
\qbezier(39.33,43.00)(41.00,41.00)(42.67,43.00)
\put(32.33,45.34){\makebox(0,0)[cc]{$\frac{n\pi}{2\la}$}}
\put(38.23,43.34){\makebox(0,0)[cc]{$n$}}
\put(36.89,51.78){\makebox(0,0)[cc]{$n+l$}}
\put(44.11,42.67){\makebox(0,0)[cc]{$l$}}
\bezier{12}(43.89,40.11)(44.67,41.33)(46.00,40.67)
\put(50.00,35.67){\makebox(0,0)[lc]{$\frac{w_{n_{2k}}-\nu_{n_{2k-1}+n}}{2}$}}
\bezier{24}(42.78,25.22)(43.22,22.22)(45.89,23.22)
\put(50.00,24.33){\makebox(0,0)[lc]{$w_{n_{2k}}$}}
\multiput(36.11,-1.23)(0,1){8}{\circle*{0.25}}
\bezier{8}(35.22,4.00)(35.45,5.44)(36.11,4.66)
\put(50.00,3.55){\makebox(0,0)[lc]{$\frac{\nu_{n_{2k-1}}-w_{n_{2k}}}{2}$}}
\put(30.00,3.55){\makebox(0,0)[cc]{$n+l$}}
\put(36.00,45.33){\vector(4,-1){5.11}}
\put(50.00,30.00){\makebox(0,0)[lc]{$\st{n_{1},\ldots,n_{2k-2}}$}}
\put(50.00,51.94){\makebox(0,0)[lc]{$u=\nu_{n-n_{2k}}$}}
\color{black}
\end{picture}
\ca{As \ref{fig:fb3}, outer legs replaced by all-breather version}
\label{fig:os1} }
\end{center}
\end{figure}

\pagebreak


\begin{thebibliography}{1}

\bibitem{GhoshZam}
\rauthor{S. Ghoshal and A. Zamolodchikov}
\rname{Boundary S matrix and boundary state in two-dimensional integrable
quantum field theory}
\journal{Int. J. Mod. Phys. }{A9}{1994}{3841--3885}%
\preprint{hep-th/9306002}

\bibitem{Kane}
\rauthor{C. Kane and M. Fisher}
\rname{Transmission through barriers and resonant tunneling in an
interacting one-dimensional electron gas}
\journal{Phys. Rev.}{B46}{1992}{15233--15262}%

\bibitem{Wen}
\rauthor{X.G. Wen}
\rname{Chiral Luttinger liquid and the edge excitations in the
fractional quantum Hall states}
\journal{Phys. Rev.}{B41}{1990}{12838--12844}%

\bibitem{Fendley}
\rauthor{P. Fendley, A.W.W. Ludwig and H. Saleur}
\rname{Exact Conductance through Point Contacts in the $\nu=1/3$
Fractional Quantum Hall Effect}
\journal{Phys. Rev. Lett.}{74}{1995}{3005--3008}%
\preprint{cond-mat/9408068}

\bibitem{Saleur}
\rauthor{H. Saleur}
\rname{Lectures on non perturbative field theory and quantum
impurity problems}
in the proceedings of the 1998 Les Houches Summer School%
\preprint{cond-mat/9812110}\\
\rauthor{H. Saleur}
\rname{Lectures on non perturbative field theory and quantum
impurity problems. Part II}%
\xpreprint{cond-mat/0007309}

\bibitem{Ghoshal}
\rauthor{S. Ghoshal}
\rname{Bound State Boundary S-Matrix of the sine-Gordon Model}
\journal{Int. J. Mod. Phys.}{A9}{1994}{4801--4810}%
\preprint{hep-th/9310188}

\bibitem{Skorik}
\rauthor{S. Skorik and H. Saleur}
\rname{Boundary bound states and boundary bootstrap in the sine-Gordon
model with Dirichlet boundary conditions}
\journal{J. Phys.}{A28}{1995}{6605--6622}%
\preprint{hep-th/9502011}

\bibitem{DTW}
\rauthor{P. Dorey, R. Tateo and G.M.T. Watts}
\rname{Generalisations of the Coleman-Thun mechanism and boundary
reflection factors}
\journal{Phys. Lett.}{B448}{1999}{249--256}%
\preprint{hep-th/9810098}

\bibitem{Zsquare}
\rauthor{A.B. Zamolodchikov and Al.B. Zamolodchikov}
\rname{Factorized S-matrices in two dimensions as the exact
solutions of certain relativistic quantum field theory models}
\journal{Ann. Phys.}{120}{1979}{253--291}%

\bibitem{Faddeev}
\rauthor{L. Takhtadjian and L. Faddeev}
\journal{Theor. Math. Phys.}{21}{1974}{160}\\
\rauthor{V. Korepin and L. Faddeev}
\journal{Theor. Math. Phys.}{25}{1975}{147}

\bibitem{Pillin}
\rauthor{M. Pillin}
\rname{Exact two-particle matrix elements in S-matrix preserving
deformation of integrable QFTs}
\journal{Phys. Lett.}{B448}{1999}{227-233}%
\preprint{hep-th/9812106}

\bibitem{Barnes}
\rauthor{E.W. Barnes}
\rname{The Theory of the Double Gamma Function}
\journal{Phil. Trans. Roy. Soc.}{A196}{1901}{265--387}
\rname{On the Theory of the Double Gamma Function}
\journal{Trans. Cambridge Phil. Soc.}{19}{1904}{376--425}

\bibitem{Jimbo}
\rauthor{M. Jimbo and T. Miwa}
\rname{QKZ equation with $|q|=1$ and correlation functions of the XXZ model 
in the gapless regime}
\journal{J. Phys.}{A29}{1996}{2923--2958}%
\preprint{hep-th/9601135}

\bibitem{FringK}
\rauthor{A. Fring and R. K\"{o}berle}
\rname{Factorized Scattering in the Presence of Reflecting Boundaries}
\journal{Nucl. Phys.}{B421}{1994}{159--172}%
\preprint{hep-th/9304141}

\bibitem{CT}
\rauthor{S. Coleman and H.J. Thun}
\rname{On the Prosaic Origin of the Double Poles in the Sine-Gordon
S-matrix}
\journal{Comm. Math. Phys.}{61}{1978}{31}

\bibitem{CM}
\rauthor{P. Christe and G. Mussardo}
\rname{Elastic S-matrices in (1+1) dimensions and Toda field theories}
\journal{Int. J. Mod. Phys.}{A5}{1990}{4581--4627}

\bibitem{BCDS}
\rauthor{H.W. Braden, E. Corrigan, P.E. Dorey and R. Sasaki}
\rname{Affine Toda field theory and exact S-matrices}
\journal{Nucl. Phys.}{B338}{1990}{689--746}

\bibitem{BCDSpole}
\rauthor{H.W. Braden, E. Corrigan, P.E. Dorey and R. Sasaki}
\rname{Multiple poles and other features of Affine Toda Field Theory}
\journal{Nucl. Phys.}{B356}{1991}{469--498}%

\bibitem{del}
\rauthor{G.W. Delius, M.T. Grisaru, and D. Zanon}
\rname{Exact S-matrices for Nonsimply-Laced Affine Toda Theories}
\journal{Nucl. Phys.}{B382}{1992}{365--408}%
\preprint{hep-th/9201067}

\bibitem{CDS}
\rauthor{E. Corrigan, P.E. Dorey and R. Sasaki}
\rname{On a Generalised Bootstrap Principle}
\journal{Nucl. Phys.}{B408}{1993}{579--599}%
\preprint{hep-th/9304065}

\bibitem{pedrev}
\rauthor{P. Dorey}
\rname{Exact S-matrices}
in the proceedings of the 1996 E\"{o}tv\"{o}s Graduate School,
\xpreprint{hep-th/9810026}

\bibitem{DPTW}
\rauthor{P. Dorey, A. Pocklington, R. Tateo and G.M.T. Watts}
\rname{TBA and TCSA with boundaries and excited states}
\journal{Nucl. Phys.}{B525}{1998}{641--663}%
\preprint{hep-th/9712197}

\bibitem{DG}
\rauthor{G.W. Delius and G.M. Gandenberger}
\rname{Particle reflection amplitudes in $a^{(1)}_n$ Toda field theories}
\journal{Nucl. Phys.}{B554}{1999}{325--364}%
\preprint{hep-th/9904002}

\end{thebibliography}
\end{document}